\documentclass[twocolumn]{aastex63}
\usepackage{amsmath}
\usepackage{lineno}
\usepackage{ulem}

\newcommand{\Acc}{A_{\rm cc}}
\newcommand{\AIa}{A_{\rm Ia}} 
\newcommand{\RIa}{R_{\rm Ia}^X}
\newcommand{\aIa}{\alpha_{\rm Ia}}
\newcommand{\acc}{\alpha_{\rm cc}}

\newcommand{\xmg}{[{\rm X}/{\rm Mg}]} 
 
\newcommand{\mgh}{[{\rm Mg}/{\rm H}]}
\newcommand{\feh}[0]{[{\rm Fe/H}]} 
\newcommand{\pIa}{p_{\rm Ia}^{\rm X}}
\newcommand{\pcc}{p_{\rm cc}^{\rm X}}
\newcommand{\pIasun}{p_{\rm Ia, \odot}^{\rm X}}
\newcommand{\pccsun}{p_{\rm cc, \odot}^{\rm X}}
\newcommand{\fcc}{f_{\rm cc}}

\newcommand{\Msun}{\rm M_{\odot}}
\newcommand{\MZAMS}{\rm M_{\rm ZAMS}}

\defcitealias{griffith19}{GJW}
\defcitealias{sukhbold2016}{S16}
\defcitealias{Kroupa1993}{KTG93}
\defcitealias{lc18}{LC18}

\newcommand{\add}[1]{\textcolor{black}{#1}}

\received{2021 March 17}
\revised{2021 August 3}
\accepted{2021 August 5}

\shorttitle{BH Formation and CCSN Yields}
\shortauthors{Griffith et al.}

\graphicspath{{./}{}}
\begin{document}

\title{The Impact of Black Hole Formation on Population Averaged Supernova Yields}

\correspondingauthor{Emily Griffith}
\email{griffith.802@osu.edu}

\author[0000-0001-9345-9977]{Emily J. Griffith}
\affiliation{The Department of Astronomy and Center of Cosmology and AstroParticle Physics, The Ohio State University, Columbus, OH 43210, USA}

\author[0000-0002-1728-1561]{Tuguldur Sukhbold}
\affiliation{The Department of Astronomy and Center of Cosmology and AstroParticle Physics, The Ohio State University, Columbus, OH 43210, USA}

\author[0000-0001-7775-7261]{David H. Weinberg}
\affiliation{The Department of Astronomy and Center of Cosmology and AstroParticle Physics, The Ohio State University, Columbus, OH 43210, USA}
\affiliation{The Institute for Advanced Study, Princeton, NJ, 08540, USA}

\author[0000-0001-7258-1834]{Jennifer A. Johnson}
\affiliation{The Department of Astronomy and Center of Cosmology and AstroParticle Physics, The Ohio State University, Columbus, OH 43210, USA}

\author[0000-0002-6534-8783]{James W. Johnson}
\affiliation{The Department of Astronomy and Center of Cosmology and AstroParticle Physics, The Ohio State University, Columbus, OH 43210, USA}

\author[0000-0002-0743-9994]{Fiorenzo Vincenzo}
\affiliation{The Department of Astronomy and Center of Cosmology and AstroParticle Physics, The Ohio State University, Columbus, OH 43210, USA}

\begin{abstract}

The landscape of black hole (BH) formation -- which massive stars explode as core collapse supernovae (CCSN) and which implode to BHs -- profoundly affects the \add{initial mass function}-averaged nucleosynthetic yields of a stellar population.  Building on the work of \cite{sukhbold2016}, we compute IMF-averaged yields at solar metallicity for a wide range of assumptions, including neutrino-driven engine models with extensive BH formation, models with a simple mass threshold for BH formation, and a model in which all stars from $8-120\Msun$ explode. For plausible choices, the overall yields of $\alpha$-elements span a factor of three, but changes in relative yields are more subtle, typically $0.05-0.2$ dex. For constraining the overall level of BH formation, ratios of C and N to O or Mg are promising diagnostics.  For distinguishing complex, theoretically motivated landscapes from simple mass thresholds, abundance ratios involving Mn or Ni are promising because of their sensitivity to the core structure of the CCSN progenitors.  We confirm previous findings of a substantial (factor $2.5-4$) discrepancy between predicted O/Mg yield ratios and observationally inferred values, implying that models either overproduce O or underproduce Mg. No landscape choice achieves across-the-board agreement with observed abundance ratios; the discrepancies offer empirical clues to aspects of massive star evolution or explosion physics still missing from the models.  We find qualitatively similar results using the massive star yields of \cite{lc18}.  We provide tables of IMF-integrated yields for several landscape scenarios, and more flexible user-designed models can be implemented through the publicly available {\tt Versatile Integrator for Chemical Evolution} ({\tt VICE}; https://pypi.org/project/vice/).

\end{abstract}


\section{Introduction} \label{sec:intro}

To explode, or not to explode? That is the question at the end of every massive star’s life. Its death, through Fe-core collapse, can result in a successful explosion with associated nucleosynthesis and neutron star formation, a failed explosion/implosion with production of a black hole (BH), or a successful explosion with subsequent fallback that triggers BH formation \add{\citep[e.g.][]{heger2003, bernhard2020}}.  Empirical results support the existence of all three pathways, as it is clear that some stars explode as core collapse supernovae (CCSN) and leave behind a neutron star, while it is also clear that stellar mass BHs exist \citep[eg.][]{remillard2006}. Both theoretical models and empirical arguments imply that CCSN are the primary source of most $\alpha$-elements and make substantial contributions to Fe-peak and weak $s$-process elements in stars with solar abundances. In this paper, we examine the impact of BH formation on the population-averaged yields of elements from He and C up to Y, Zr, and Nb. We aim to provide useful inputs to models of galactic chemical evolution (GCE) and to identify abundance diagnostics for the physics of BH formation.

While some models of CCSN yields have assumed a simple cut in zero-age main sequence mass ($\MZAMS$) to separate explosions and BH formation \citep[e.g.,][]{lc18}, theoretical studies suggest that the ``landscape'' of BH formation is complex, showing regions of explosions interleaved with regions of collapse \citep[e.g.][]{ugliano2012, pejcha2015, sukhbold2016, ertl2016}. The complex nature of the explodability landscape emerges from intricacies of shell evolution in massive progenitors, such as the location and timing of the C and O burning shell \citep[e.g.][]{sukhbold2014}, which affects density and entropy structures of pre-collapse cores. Explosions of stars with $\MZAMS>40\Msun$ are sensitive to mass loss and its effect on compactness of the stellar core. The works cited above and the landscapes that we will explore in this paper report a binary `yes’ or `no’ to BH production for a given $\MZAMS$ and metallicity. The true BH landscape may be even more complex, with each star having a probability of explosion that depends in detail on its composition, binarity, mass loss, and other properties \citep{clausen2015}. The binary landscapes shown here should be taken as representing the most likely explosion outcome at a given $\MZAMS$. 

We base our yield predictions on the solar metallicity supernova models of \citet[][hereafter S16]{sukhbold2016}. A similar study of yields at higher and lower metallicities would be interesting as well, but we focus on solar $Z$ due to the availability of quality stellar models and yields. \add{While massive star explodability has been modeled for other metallicity ranges \citep[e.g.][]{pejcha2015}, stellar yields for complex BH landscapes have not yet been calculated at non-solar $Z$.} \citetalias{sukhbold2016} consider a range of ``engines,’’ corresponding to differently calibrated neutrino transport explosion models. Relative to models that adopt a simple arbitrary mass cuts and explosion energies \citep[e.g.][]{woosley1995}, adopting calibrated neutrino driven central engines gives a more physically informed view of CCSN and the BH landscape \citep{oconnor2011, ugliano2012, horiuchi2014, sukhbold2014, pejcha2015, ertl2016, muller2016, sukhbold2016, ebinger2019, mabanta2019, ertl2020}.
But BH landscape models pose their own uncertainties including the fact that they are not truly ab-initio, its reliance on arbitrary calibrations (often to SN 1987A), and its highly approximate treatments of multi--dimensional effects. In addition, the current theoretical understanding is only weakly supported by observations \citep[e.g., direct progenitor imaging,][]{smartt2015,adams2017}.

We explore how more or less explosive models impact elemental ratios and present the yield differences between landscapes with islands of explodability and those with upper mass cutoffs. Our tabulated yield predictions can be adopted in GCE models to represent a variety of assumptions about BH formation. These yields and their consequent abundance ratios can also be used as an observational diagnostic of the Milky Way’s BH landscape. If the IMF\footnote{IMF = Initial Mass Function}-averaged yields of two elements have different dependence on explodability, then their ratio would differ between landscapes. We aim to determine which elements might have the most variation with BH landscape, and which elemental ratios would be the best diagnostics of BH formation. We strive to find abundance ratios that could separate landscapes with islands of explodability from those with a simple maximum explosion mass, as well as abundance ratios that could distinguish between the more and less explosive scenarios.

Uncertainties in CCSN yields are an ongoing issue in GCE models, as empirical yield estimates do not agree with CCSN yields for some elements. One such example is the overproduction of O or underproduction of Mg. The dominant production of these two elements in CCSN \citep{andrews2017} suggests that CCSN yields should reproduce the solar O/Mg ratio, but many solar metallicity studies find Mg underproduction \citep{sukhbold2016, rybizki2017, lc18}. We explore the severity and implications of CCSN yield discrepancies with empirical results and discuss the ability of IMF changes or more/less explosive BH landscapes to resolve the inconsistencies.

CCSN yields are traditionally compared to and evaluated against the solar mixture. The sun, however, holds material produced from a variety of processes, including CCSN, Type-Ia supernovae (SNIa), and asymptotic giant branch (AGB) stars. \citet{weinberg2019} introduced an empirical ``two-process'' model that isolates the CCSN and SNIa contributions to elements in Milky Way disk stars by fitting the median abundance trends of high-[Mg/Fe] and low-[Mg/Fe] stellar populations. We compare our theoretical yield predictions to the CCSN abundance patterns inferred from these empirical decompositions by \citet{weinberg2019} and \citet{griffith19}, based respectively on data from the APOGEE\footnote{APOGEE = Apache Point Observatory Galactic Evolution Experiment, conducted as a part of the Sloan Digital Sky Survey III \citep[SDSS-III][]{eisenstein2011} and IV \citep[][]{blanton2017}} \citep{majewski17} and GALAH\footnote{GALactic Archaeology with HERMES} \citep{desilva2015, martell2017} surveys. Although these decompositions have uncertainties due to the simplicity of the model and systematic uncertainties in the APOGEE and GALAH abundances, they are clearly a step in the desired direction, allowing a more accurate assessment of the models’ success and failures than a simple comparison to solar abundances without accounting for SNIa or AGB contributions.

We begin with a more detailed discussion of our CCSN models in Section~\ref{sec:ccsn_yields}, outlining the construction of a new, fully exploding yield set. We explore the dependence of elemental yields on $\MZAMS$ in Section~\ref{sec:yield_v_props} and discuss the power of pre-SN properties in predicting successful explosions. In Section~\ref{sec:IMF_int} we present the IMF-averaged abundances for all of our explosion landscapes and compare their relative abundance ratios in Section~\ref{sec:ab_ratios}. Here we discuss abundance ratios that may be diagnostic of the BH landscape, compare our results with empirical data, and explore the differences between the complex and simple BH landscapes. Section~\ref{sec:discussion} discusses sources of uncertainty in our predictions, implications of our results for supernova physics, and the impact of choice of IMF. We summarize our results in Section~\ref{sec:summary}.

\section{CCSN Yields}\label{sec:ccsn_yields}

In this paper, we explore the CCSN yields based on 1-dimensional calibrated neutrino--driven explosion models from \citetalias{sukhbold2016}. The 200 pre-SN stars used in \citetalias{sukhbold2016} were non-rotating solar metallicity models computed with the implicit hydrodynamics code KEPLER \citep{weaver1978}, ranging in birth mass between $9-120\ \Msun$. Lower and intermediate mass ($\MZAMS < 30 \Msun$) models were from \citet{woosley2015} and \citet{sukhbold2014}, and higher mass models were utilized from \citet{woosley2007}. The evolution from the Fe-core collapse through core bounce and post-bounce accretion were followed with P-HOTB \citep[for details see also][]{ertl2016,ugliano2012}, and the nucleosynthesis of stars that produced a successful explosion were computed with KEPLER by mimicking the results of P-HOTB.

\subsection{Engine-driven Explosions}\label{subsec:engines}

\begin{figure*}[!htb]
\begin{centering}
 \includegraphics[width=\textwidth]{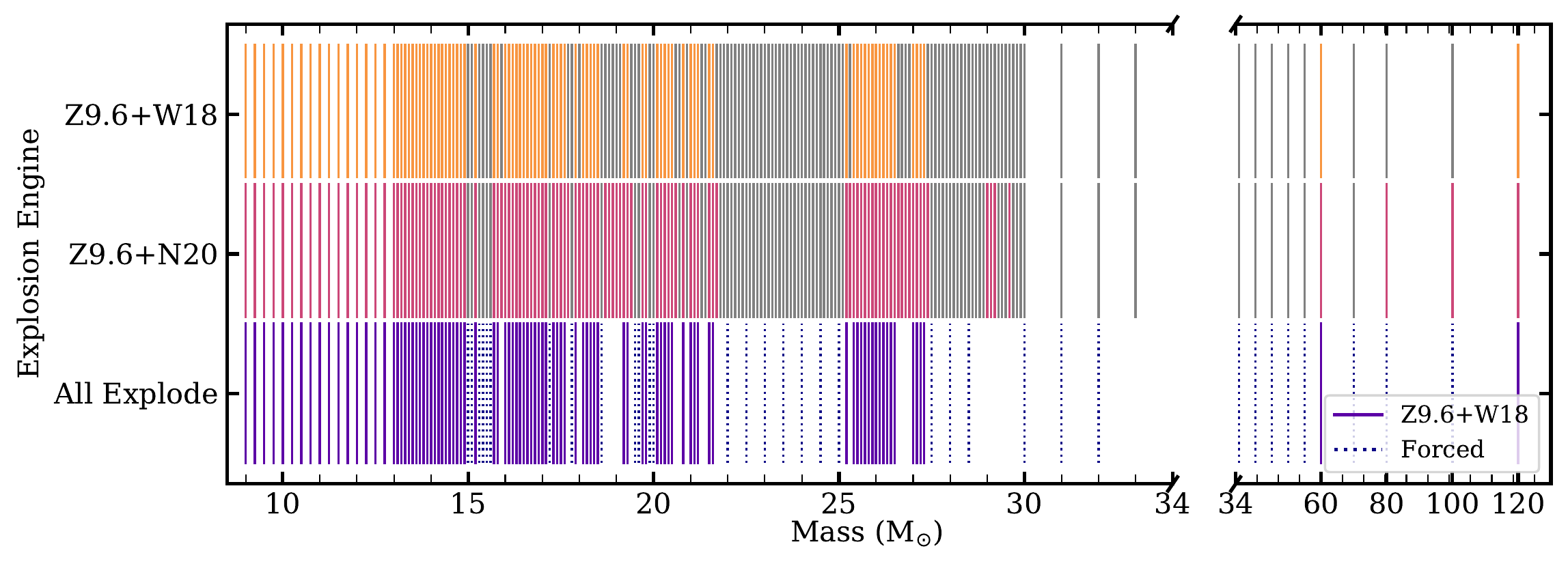}
 \caption{Z9.6+W18 (top, orange), Z9.6+N20 (middle, pink), and All Explode (bottom, purple \& blue) landscapes. These maps show the progenitor masses which explode as CCSN under each explosion engine in color and those which collapse in grey. All Explode shows the Z9.6+W18 explosions in purple and the masses where we force explosions as dotted blue lines. We have changed the x-axis scale at 34$\Msun$ for ease of viewing.}
 \label{fig:landscape}
 \end{centering}
\end{figure*}

The explosion ``engines'' of \citetalias{sukhbold2016} were based on five different calibrations to SN 1987A (S19.8, N20, W15, W18, W20), used for stars with masses $\geq 12\Msun$, and a single calibration to SN 1054 (Z9.6) that is used for lighter stars. The Z9.6 engine is used in conjunction with the SN 1987A-calibrated engines to obtain mass coverage from $9-120\ \Msun$. The final outcomes for each pre-SN star, i.e., whether the star explodes or not and the properties of the explosion in successful SNe, are uniquely tied to the pre-SN core structure of the progenitor. We refer to the collective outcomes on the mass-space as the explodability landscape or as the BH formation landscape. The successful explosions from Z9.6+W18 and Z9.6+N20 combinations are illustrated as vertical bars in Figure~\ref{fig:landscape}, similar to Figure 13 in \citetalias{sukhbold2016}. 

The published yields from \citetalias{sukhbold2016} include all species up to $^{209}$Bi, and cover Z9.6+W18 and Z9.6+N20 results. In each successful explosion the yields include contributions from the ejected SN matter and mass loss during hydrostatic stellar evolution, which we refer to generically as winds. The yields for stars that failed to explode include the contribution from winds alone. While a significant fraction or all of the remaining stellar envelope could be ejected during the implosion \citep[e.g.,][]{lovegrove2013}, it has a negligible input to the yields and is ignored in this study for simplicity.

The yield contributions from supernova ejecta are all computed at 200 seconds after the Fe-core collapse, when all of the explosive nucleosynthesis has ended. To account for some of the subsequent decay of radioactive isotopes we convert all $^{26}$Al to $^{26}$Mg and all $^{56}$Ni to $^{56}$Fe. While other radioactive species introduce small changes to the final yield, we only force the decay of $^{26}$Al and $^{56}$Ni as we will discuss Fe and Mg in detail within this paper.

\subsection{Creating a Fully Exploding Yield Set}\label{subsec:all_exp}

To conduct the desired comparison between BH landscapes of varying degrees of explodability, we require explosion results and their corresponding sets of yields for each case. Ideally this should be done by performing a diverse range of calibrations to the neutrino-powered engine and by following the nucleosynthesis in each successful explosion. However, in this paper we pursue a much simpler approach utilizing the existing models of \citetalias{sukhbold2016} and without creating any new neutrino-hydrodynamics calculations. Based on the properties of Z9.6+W18 results from \citetalias{sukhbold2016} we construct a baseline set of yields in which all stars die in a supernova. This is achieved by effectively forcing all stars that were deemed to implode by the W18 engine to explode instead. All new models were computed using KEPLER with its piston scheme and large nucleosynthesis network. 

We compute these artificial explosions (shown as dotted lines in Figure~\ref{fig:landscape}) in a way that somewhat mimics the characteristics of successful explosions of the W18 engine. In particular, we observe the unique correlations between the properties in the presupernova core structure of progenitor stars and their key explosion properties (e.g., final mass separation, kinetic energy of the explosion). For instance, as had been noted before \citep[e.g.,][]{sukhbold2018,woosley2002}, the final mass separation in neutrino-driven explosion simulations closely tracks the so-called $M_4$ point in the core, the mass coordinate where the entropy per baryon exceeds 4$/k_{\rm B}$ going outward, often found at the base of the last O burning shell. In addition, the time of the bounce and the final explosion energies are tightly correlated with the radial gradient of mass at the location of $M_4$, often referred to as $\mu_4$. The bounce is delayed and generally leads to lower kinetic energies when the entropy jump is steeper. 

These correlations were used to construct the motion of the piston in our artificial explosions. The final mass separation is directly set by the Lagrangian coordinate of $M_4$, while bounce time and the final kinetic energies are determined through a simple linear fit based on their relations to $\mu_4$. The inward parabolic collapse of the piston is governed by the mass cut and the radial (taken as a constant 135 km) and temporal coordinates of the bounce. The outward motion from the bounce point until $10^9$ cm is obtained by iterating on the gravitational potential until the final kinetic energy of the ejected material is obtained (for details see section 3.2 of \citetalias{sukhbold2016}). Once the hydrodynamics converge, we extend the calculation to 200 seconds using a large nucleosynthesis network.

Our fully exploding yield set, which we refer to as `All Explode' in the rest of the paper, is created by compiling together the Z9.6+W18 yields from \citetalias{sukhbold2016} with these forced artificial explosion results. This baseline set allows us to construct yields based on any desired explosion landscape. As a simple test of this concept, we take a set of yields based on a different engine combination (Z9.6+N20) and compare it with yields we obtain from our baseline `All Explode' set after we impose the Z9.6+N20 explosion landscape (see Figure~\ref{fig:landscape}). We find almost identical results, suggesting that the yields estimated from an interpolation of artificial explosions, described above, are a reasonable representation of the explosive yields these stars would have had, if they did explode. 

\section{Explosive \& Wind Yield Dependences on Progenitor Properties} \label{sec:yield_v_props}

\subsection{Yield \& Progenitor Mass} \label{subsec:yield_v_mass}

\begin{figure*}[!htb]
\begin{centering} 
 \includegraphics[width=.8\textwidth]{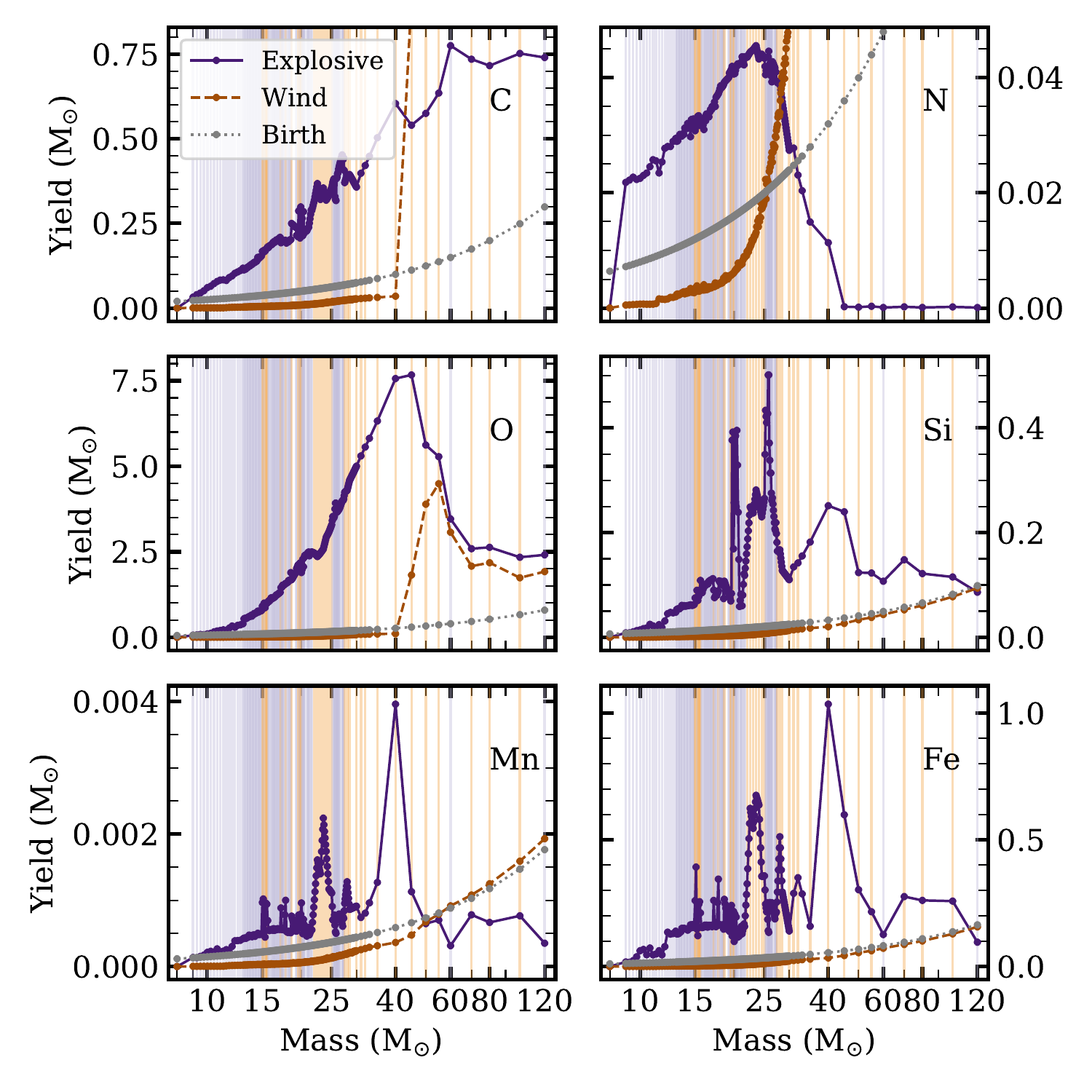} 
 \caption{Explosive (dark purple, solid line) and wind (dark orange, dashed line) yields in $\Msun$ produced per star of a given progenitor mass for C (top left), N (top right), O (middle left), Si (middle right), Mn (bottom left), and Fe (bottom right). The elemental birth abundances per star are also plotted as the grey dotted line. Background colored lines indicate successful explosions under the Z9.6+W18 engine (light purple) and forced explosions (light orange). The wind yields of C and N run off the upper end of the plot, but continue on their trajectory for high masses, with values of $6.562\Msun$ (C) and $0.196\Msun$ (N) at $\MZAMS = 60 \Msun$.}
 \label{fig:mass_yield}
\end{centering}
\end{figure*}

The IMF-averaged yield of some element, X, produced by a CCSN model is dependent upon the BH landscape, as stars of different birth masses may or may not contribute to it depending on their final outcome. Even if all stars did explode, not all elemental yields follow the same mass dependence. Figure~\ref{fig:mass_yield} plots the final explosive yields along with wind contributions and birth abundances for C, N, O, Si, Mn, and Fe for our All Explode CCSN model. By ``explosive yield'' we refer to the final yield in the ejected material after the CCSN explosion, and the ``wind yield'' refers to elemental mass lost prior to the supernova during the star's hydrostatic evolution. For a given element, if the star does not experience significant production or enrichment (e.g., due to mixing) in its outer layers then the wind yield is nearly equal to the birth abundance. Stars that collapse to BH without a supernova will only release their wind yields to the ISM but not their explosive yields.

We clearly see that the yield of each element has a unique mass dependence. The explosive C yield monotonically increases across all mass ranges, while the wind yields jump rapidly at 40$\Msun$ and become the dominant source of C production at all higher masses. Wind yields also dominate the production of N above $~30\Msun$, though they increase more gradually than C. Explosive N yields steadily increase, then turn over at around $23\Msun$, the birth mass at which the presupernova star retains the highest mass H-envelope. Because N is the bottleneck of the primary CNO-cycle, its explosive yield is tightly correlated with the envelope mass of the presupernova star.

The explosive O yield, substantially contributed by the He and C burning shells, mirrors C for stars of progenitor mass $9-40 \Msun$, then decreases for stars of higher mass. Due to its tight correlation in this mass range, O is often utilized as a proxy for estimating the birth mass (or the He-core mass)\footnote{Following standard usage, the term ``He-core mass'' refers to the central zone comprised of elements heavier than H, including He, C, O, etc.} of Type II supernova progenitors \citep[e.g.,][]{jerkstrand2015}. The wind yield remains comparable to the birth abundance until $\sim 40 \Msun$, but it then increases to nearly equal the explosive yields for stars with masses greater than $40 \Msun$. The yield behavior of C and O changes around $40 \Msun$, as the pre-SN models from \citetalias{sukhbold2016} begin to shed their He and C shells in winds for $\MZAMS \gtrsim 45\Msun$. As the He and C shells are lost, explosive yields from He and C diminish, and correspondingly the wind yields of C, N, and O increase. The exact $\MZAMS$ where stars completely lose their H-envelope, and how much of the He and C shells gets stripped away at higher masses, is not precisely known. Envelope and shell retention depend sensitively on the red supergiant and Wolf-Rayet mass loss rates, both of which are poorly constrained \citep[e.g.,][]{beasor2018,yoon2017}, remaining one of the key uncertainties of massive star evolution models.

The explosive yields of Si, Mn, and Fe, elements made deeper in the star or ejecta, behave differently and have more variation than O, C, and N. Explosive Si yields have broad peaks at 23 and $40 \Msun$. The sharp peaks around $20 \Msun$ and $26 \Msun$ correspond to models where O and C burning shells have merged in the final few years of evolution \citep[e.g.,][]{sukhbold2014}. Wind yield contributions to Si remain small regardless of progenitor mass; the net yields after subtracting birth abundances are nearly zero. 

The explosive yields of Mn and Fe also peak at around $23 \Msun$ and $40 \Msun$, and their wind yields are comparable to birth abundances for both elements at all masses, indicating no net wind production. The explosive yields of Mn and Fe (also Si) tightly correspond to core structure of pre-SN models, which plays an important role in determining properties of the explosion and the nucleosynthesis of Fe-group species. The lightest massive stars have highly compact structures where the density drops steeply outside their small Fe cores, which leads to lower energy explosions and smaller production of Fe-group species. These stars are also easier to blow up because the amount of ram pressure that needs to be overcome to produce an explosion is smaller due to rapidly declining density. The opposite is generally true for higher mass stars, due to the interplay of convective burning episodes in their cores during the final few thousand years. This causes final structures that are highly non-monotonic as a function of birth mass \citep{sukhbold2014,sukhbold2018}. The variations seen in the yields of Si, Mn, and Fe (Figure~\ref{fig:mass_yield}) are largely driven by the changes in the structure of the pre-SN stellar core.

\begin{figure}[!htb]
\begin{centering}
 \includegraphics[width=\columnwidth]{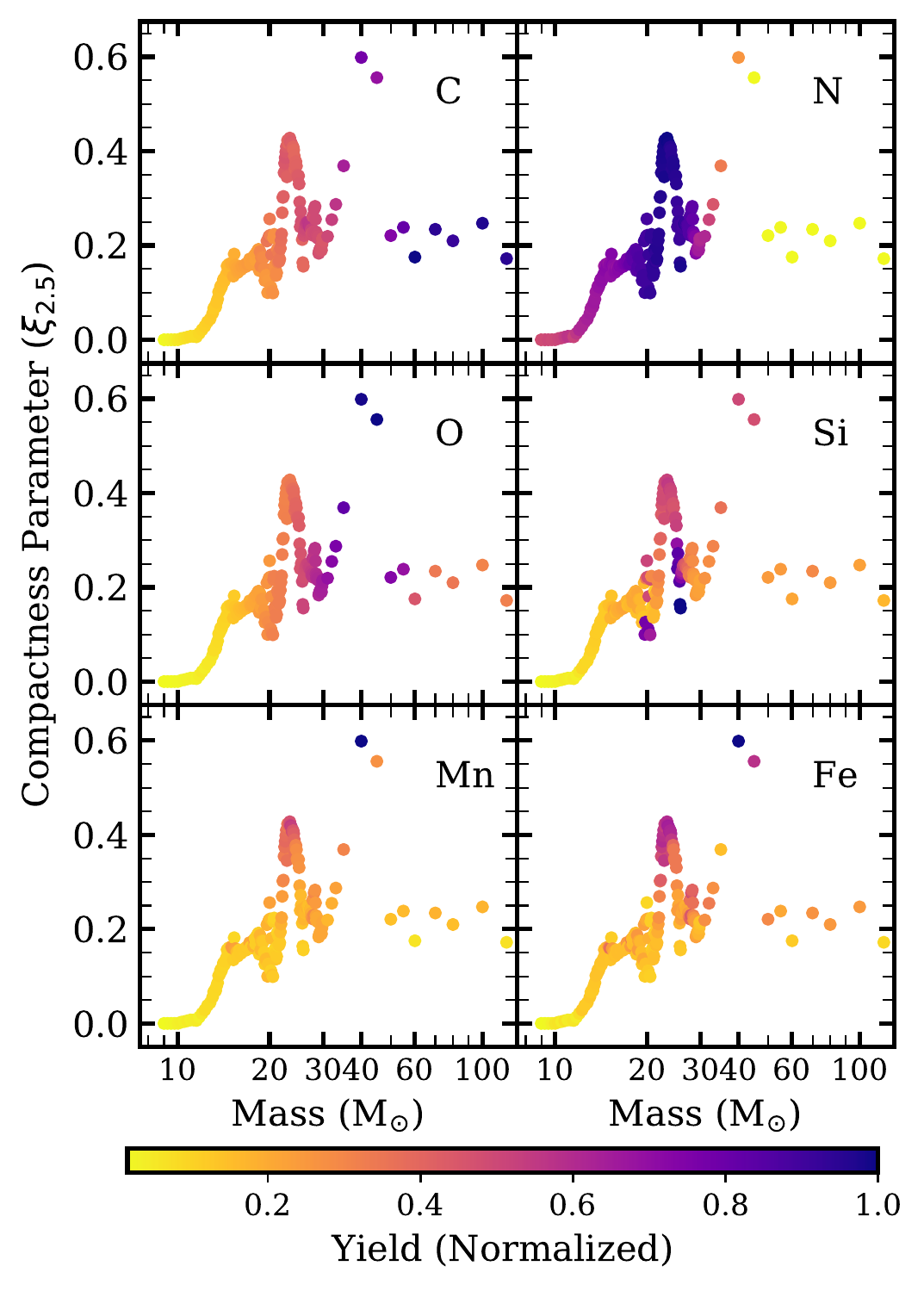}
 \caption{Compactness parameter ($\xi_{2.5}$) as a function of stellar progenitor mass, colored by the normalized elemental explosive yield (omitting wind contributions) of C (top left), N (top right), O (middle left), Si (middle right), Mn (bottom left), and Fe (bottom right). The points are the same in each panel but the color gradient changes. Elements whose yield increases primarily with mass (e.g. O) show a gradient left to right. Elements whose yield increases strongly with compactness parameter (e.g. Fe) show a gradient from bottom to top.}
 \label{fig:mass_cp}
 \end{centering}
\end{figure}

One simple way of characterizing this final structure is the compactness parameter ($\xi_{2.5}$), which measures the inverse of the radius enclosing the innermost $2.5 \Msun$ in the pre-SN star \citep{oconnor2011}. The correlation of yield with compactness parameter can be seen in Figure~\ref{fig:mass_cp}, which plots $\xi_{2.5}$ vs. $\MZAMS$ with points colored by the normalized explosive yield for C, N, O, Si, Mn, and Fe. Smaller $\xi_{2.5}$ corresponds to stars with steeply dropping density outside the Fe core that are easier to explode, and higher values represent stars with dense extended cores. Many of the highest $\xi_{2.5}$ models do not die in a supernova under the Z9.6+W18 engine, but if they are forced to blow up, as we explore in this study, they create powerful explosions and produce substantial amounts of Mn, Fe, and other Fe-peak elements. In Figure~\ref{fig:mass_cp}, the elements whose explosive yield correlates with $\xi_{2.5}$, such as Mn and Fe, show a color gradient from bottom to top. This figure also illustrates $\MZAMS$-yield correlations, as elements whose explosive yields increase with progenitor mass have a color gradient progressing from left to right (e.g. C and O).
        
While we only include the yield vs. mass trends for six elements in this section, similar plots for all other elements included in \citetalias{sukhbold2016} can be found in Appendix~\ref{ap:yields} (Figure~\ref{fig:mass_yield_all}). The trends of other elements can largely be categorized by their likeness to four of the elements discussed above:

\textbf{\textit{N-like:}} He resembles N, peaking at $23 \Msun$. It too has substantial contribution from winds above $45 \Msun$ as the outer shell begins to be stripped away.

\textbf{\textit{O-like:}} Ne, Mg, Al, Cu, Zn, and all heavier elements (Ga to Mo) trace O, though with minor wind yields above $40 \Msun$. Na and F resembles O to some extent, but they show more variation with mass. The drop in Na yields at $40 \Msun$ is likely due to the loss of the C-shell and diminishing $^{23}$Na production. Similarly, the explosive yields of the weak $s$-process elements that are formed inside the He-shell decline at this mass

\textbf{\textit{Si-like:}} elements substantially contributed by O and Ne burning such as P, S, Cl, Ar, K, Ca, and Sc resemble the general variations seen for Si. They weakly mirror the changing core structure, and they often exhibit the narrow peaks around $20 \Msun$ and $26 \Msun$ due to late shell mergers. \add{As pointed out by \citet{ritter2018}, these mergers are efficient sites for the production of odd-$Z$ species. We see these mergers rarely and only at high birth mass ($\MZAMS \geq 19$), in a rough accord with some of the prior studies \citep[e.g.][]{rauscher2002, tur2007}. However, we note that their extent and occurrence rate are highly uncertain and are sensitive to the treatment of convective physics as well as any small changes during the advanced stage of evolution.} 

\textbf{\textit{Fe-like:}} Ti, V, Cr, Co, and Ni strongly correlate with compactness parameter, like Mn and Fe. All are Fe-peak elements and produced from similar nucleosynthetic processes. 

\subsection{Predicting Explodability from pre-SN Properties}\label{subsec:e0_preds}

With the All-Explode models, we can investigate any BH landscape. However, there are characteristics of the pre-SN models that impose patterns on the stars that are likely to explode or implode, creating a physically motivated way to probe a finite range of BH landscapes. \citet{ugliano2012}, \citet{pejcha2015}, \citetalias{sukhbold2016}, and others have found that neutrino-powered CCSN produce a rather complicated landscape of explodability, with regions of explosions and regions of implosions. \citetalias{sukhbold2016} provide five possible CCSN landscapes, based on five different calibrations, each sampled finely at the lower masses and coarsely at the highest masses. In this section, we produce a wider range of explosion landscapes. Rather than conduct a slew of computationally expensive stellar evolution and explosion models, we construct artificial landscapes by utilizing the properties of pre-SN stellar cores in combination with existing results of \citetalias{sukhbold2016}.

Many works have searched for correlations between the final outcomes and stellar properties prior to collapse. Simple parameters probing the pre-SN core structure, such as compactness or the Fe-core mass, can only serve as an approximate proxy to predict some CCSN explosions \citep{oconnor2011, pejcha2015, ertl2016}. \citet{ertl2016} instead propose a two-parameter approach, leveraging the combination $M_4$ and $M_4\mu_4$, which are linked to the mass infall rate and neutrino luminosity. Based on their calibrated neutrino-driven simulations (which also underlie calculations of \citetalias{sukhbold2016}) they construct a `critical curve' for separating explosions and implosions. With these two parameters, \citet{ertl2016} are able to successfully predict the final outcomes of $97\%$ of the progenitor models across the entire birth mass range, whereas with simple proxies alone the success does not exceed $\sim$90\%. 

\begin{figure}[!htb]
\begin{centering}
 \includegraphics[width=\columnwidth]{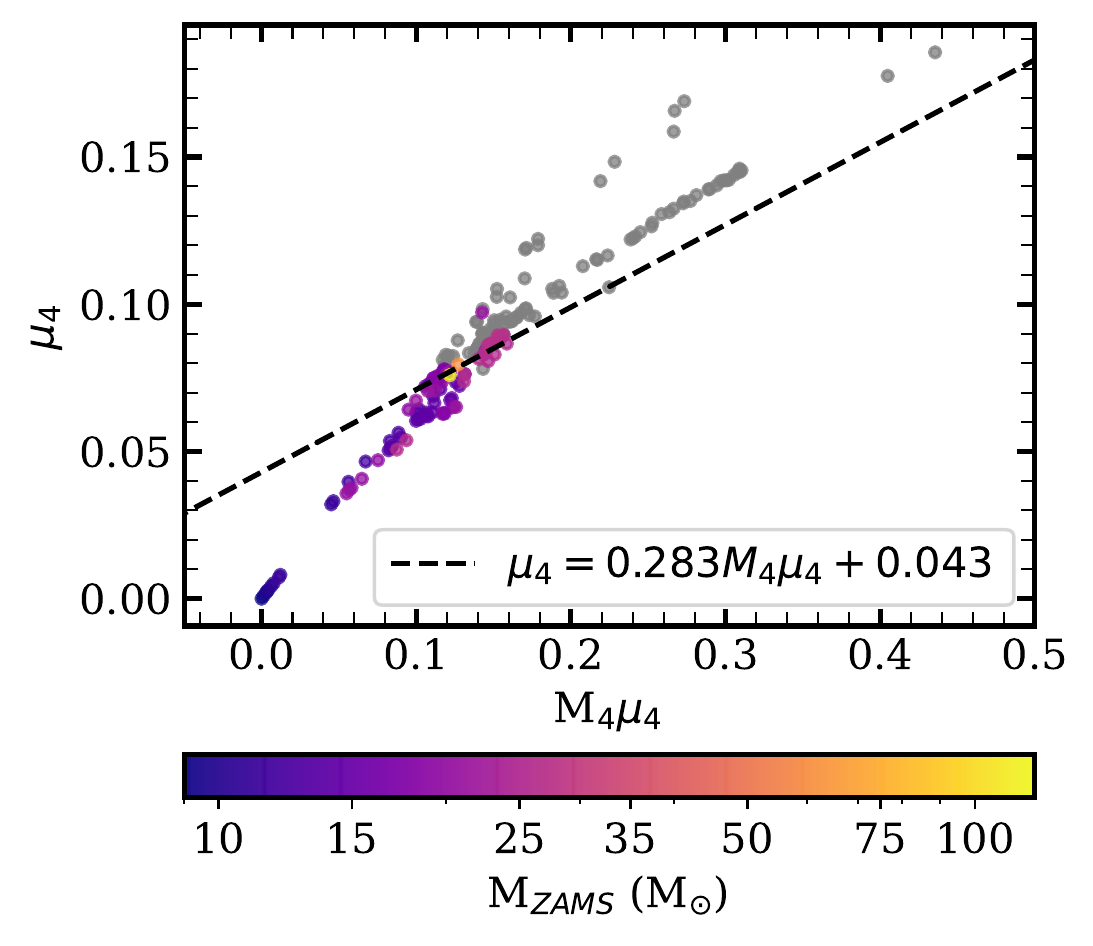}
 \caption{The $\mu_4$ vs. $M_4\mu_4$ values for progenitor stars from \citetalias{sukhbold2016}. The dividing line from \citet{ertl2016} (Equation~\ref{eq:W18_mu4}) has been overplotted as the dashed black line. \citet{ertl2016} predict that stars above this line collapse while those below explode. Successful explosions are colored by $\MZAMS$ and failed explosions are plotted in grey.}
 \label{fig:mu4}
 \end{centering}
\end{figure}

For the Z9.6+W18 engine, \citet{ertl2016} find that the exploding and non-exploding models are well separated by a line at 
\begin{equation} \label{eq:W18_mu4}
    \mu_4 = 0.283 M_4 \mu_4 + 0.043.
\end{equation}
Models with $\mu_4$ below this line explode while those above do not. We show this dividing line alongside the $\mu_4$ and $M_4\mu_4$ parameters from \citetalias{sukhbold2016} in Figure~\ref{fig:mu4}. Successful explosions are colored by $\MZAMS$ and failed explosions are in grey.

Other engines have similar slopes but different intercepts, with a larger intercept corresponding to a more energetic (and more explosive) engine. We leverage this regularity and the predictive ability of \citet{ertl2016}'s model to construct continuous BH landscapes. We take Equation~\ref{eq:W18_mu4}, round the slope to 0.28, and replace the intercept with the variable $e_0$ to define 
\begin{equation} \label{eq:e(M)}
    e(M)= 0.28 M_4 \mu_4 - \mu_4 + e_0
\end{equation}
as a function that can predict the success or failure of the explosion of a given progenitor. A progenitor of mass $M$ with $e(M)>0$ explodes and one with $e(M)<0$ does not. This function allows us to choose $e_0$, and thus the explodability of our landscape. For Z9.6+W18, $e_0$ = 0.043. With this $e_0$, the model of Equation~\ref{eq:e(M)} produces a landscape resembling the top row of Figure~\ref{fig:landscape}. A higher (lower) $e_0$ value corresponds to a more (less) energetic explosion and a more (less) explosive landscape. 

For a given $e_0$ value, we calculate $e(M)$ from the $M_4$ and $\mu_4$ parameters for all $\sim200$ pre-SN models used in \citetalias{sukhbold2016}. By interpolating between $e(M)$ values, we can determine the explodability at any $\MZAMS$, not just the Z9.6+W18 mass grid points. If $M_i$ does not explode and $M_{i+1}$ does, we can use the interpolation in $e(M)$ to find the transition mass ($M_t$) between an explosion and non-explosion, and define a continuous BH landscape.
    
\begin{figure*}[!htb]
 \includegraphics[width=\textwidth]{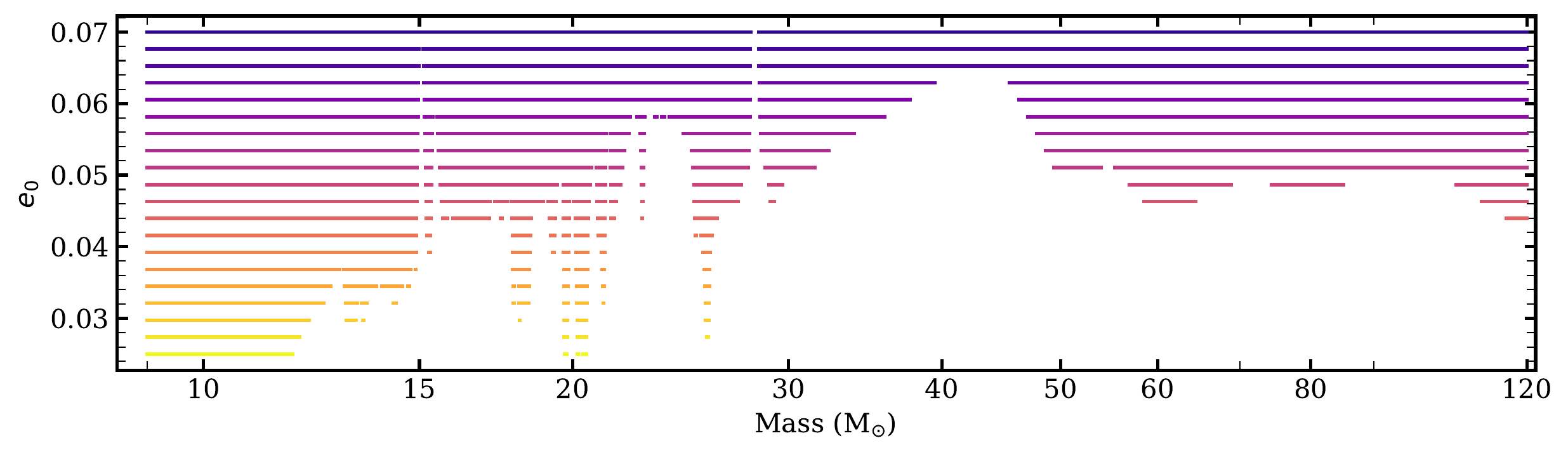}
 \caption{Continuous explosion maps as a function of progenitor mass for a range of $e_0$ values. Lines are colored by $e_0$ with lighter, yellow lines indicating low $e_0$ and darker, purple lines indicating high $e_0$. White space indicates mass regions which collapse to BHs. The Z9.6+W18 explosion model has an $e_0 =0.043$.}
 \label{fig:e0_mass}
\end{figure*}
    
Figure~\ref{fig:e0_mass} shows landscapes for 20 values of $e_0$, ranging from 0.025 to 0.07. For low values of $e_0$, the successful explosions only occur for stars of progenitor masses $\lesssim 15 \Msun$ and with narrow islands near $\sim 20$ and $26 \Msun$. These landscapes are similar to the weakest engines of \citetalias{sukhbold2016} (W20 and W15), and the explosions correspond to progenitors with the lowest compactness parameters (Figure~\ref{fig:mass_cp}). Landscapes produced from $e_0$ of $0.04-0.05$ resemble those of Z9.6+W18 and Z9.6+N20 engines with many islands of explodability. At the highest $e_0$ values, we construct landscapes that have more explosions than the most powerful one considered by \citetalias{sukhbold2016} (S19.8) and approach the All Explode landscape for the highest $e_0$. 

Our predicted continuous explosion landscapes provide a proxy for a finely sampled grid of neutrino driven CCSN explosions. While this prescription cannot capture the properties of individual explosions in any detail, it can clearly approximate the complex dependence of the final outcomes on the non-monotonically varying progenitor core structures, and on the varying power of calibrated neutrino-driven engines.
    
\section{IMF-Averaged Yields}\label{sec:IMF_int}

In this section we compute and compare the IMF-averaged yields and yield ratios predicted by the Z9.6+W18, Z9.6+N20 \citepalias{sukhbold2016}, and All Explode yield sets (Section~\ref{subsec:all_exp}) along with an upper mass cut landscape and the 20 predicted continuous explosion landscapes constructed above.

To calculate the IMF-averaged yields of the Z9.6+W18, Z9.6+N20, and All Explode yield sets, we employ the Versatile Integrator for Chemical Evolution \citep[{\tt VICE},][]{johnson2020}. {\tt VICE} allows us to easily manipulate settings in our integration. Through it we can specify the desired element, the IMF function, and the upper/lower mass bounds on star formation. With this paper, we publicly release the addition of the Z9.6+W18 and Z9.6+N20 yield sets \citepalias{sukhbold2016} as well as the All Explode yield set (Section~\ref{subsec:all_exp}) to {\tt VICE} for the reader to independently explore. We also add three new parameters to the CCSN IMF integration routine:
\begin{enumerate}
    \item \textbf{\textit{Explodability function}}: For fully exploding yield sets (i.e., those that report explosive yields for all stellar masses), the user can specify an explodability function -- a function that assigns a value from 0 to 1 to a given stellar mass indicating if it successfully explodes (1), implodes (0), or contributes some fractional value of the total explosive yields (e.g., 0.5).
    \item \textbf{\textit{Inclusion or exclusion of winds}}: We add a parameter to specify if wind yields are included in the integration. When winds are turned on, {\tt VICE} integrates the sum of explosive and wind yields. If turned off, {\tt VICE} only integrates the explosive yields. This function works for yield sets that report wind and explosive yields separately.
    \item \textbf{\textit{Net vs. gross yields}}: We add tables of birth abundance values assumed by each yield set to {\tt VICE} along with a flag to calculate net or gross yields. Net yields may return a negative value if the birth amount is more than the amount returned.
\end{enumerate}

With the addition of these parameters, the IMF-averaged \add{net} yield can be expressed as
\begin{equation} \label{eq:vice_yield}
    y_{\text{x}}^{\text{CC}} = \frac{\int_8^u (E(m) (y_{\text{x}, \text{exp}}) - m b_{\text{x}} + y_{\text{x}, \text{winds}}) \frac{dN}{dm} dm}{\int_l^u m \frac{dN}{dm} dm} ,
\end{equation}
where $E(m)$ is the explodability function, $y_{\text{x}, \text{exp}}$ is the explosive yield, $y_{\text{x}, \text{wind}}$ is the wind yield, and $b_{\text{x}}$ is the birth abundance (per $\Msun$) for a star with $\MZAMS =$ $m$.

For the results shown in this paper, we calculate IMF-averaged yields using the \citet{kroupa2001} IMF. These are net yields that include wind contributions. \add{Elemental gross yields calculated with VICE are similar to net yields for many elements and are most similar for the All Explode landscapes. The distinction between gross and net yields is insignificant for elements whose production is dominated by CCSN, as the birth abundance is a small fraction of the gross yield. Elements with significant non-CCSN sources and whose birth abundance is thus a larger fraction of their yield, such as Mn and $s$-process elements, vary the most between net and gross yields. The gross Mn yields, for example are a factor of $\sim1.5$ larger than the net yields for the All Explode landscape and a factor of $\sim3$ larger for the Z9.6+W18 landscape. Because we isolate the CCSN contribution to solar abundances in our data comparisons in Section~\ref{sec:ab_ratios}, we consider the net yield comparison to be more relevant. We note that if only considering winds, the distinction between gross and net wind yields is significant, as the wind yields and birth abundances are comparable for many elements (see Figure~\ref{fig:mass_yield_all})}

We also make frequent use of {\tt VICE}'s explodability function. For example, we produce an Explode to $40\Msun$ yield set by integrating the All Explode yields (Section~\ref{subsec:all_exp}) with the explodability function
\begin{equation}
E(x) = \begin{cases}
1 & \text{if } M < 40\Msun\\
0 & \text{if } M \geq 40\Msun 
\end{cases}
\end{equation}
to produce IMF-averaged yields for an explosion landscape with an upper explosion mass of $40\Msun$. The IMF-averaged yields for the Z9.6+W18, Z9.6+N20, Explode to 21$\Msun$, Explode to 40$\Msun$, and All Explode yield sets are given in Table~\ref{tab:yield}. Individual stellar yields for the Z9.6+W18 yield set can be found in {\tt VICE} through the look up function {\tt vice.yield.ccsne.table(element, study=`S16/W18')} or {\tt study=`S16/W18F'} for the All Explode yields.

IMF-averaged yields for the continuous explosion landscapes predicted from the linear interpolation of $M_4$ and $\mu_4$ parameters (Section~\ref{subsec:e0_preds}) are not calculated with {\tt VICE}. The continuous nature of these landscapes allow us to collapse the integral in Equation~\ref{eq:vice_yield} to a discrete sum where the trapezoid rule returns the exact result, regardless of the spacing or regularity of the grid on which $e(M)$ is tabulated, using linear interpolation to find the transition masses that separate exploding and non-exploding models for a specified $e_0$, defined by $e(M_t)=0$. In practice, this computation gives yields very similar to those computed by {\tt VICE} for the same explodability, but the exact trapezoidal sum applies for linear interpolation of the product of IMF and yield, while {\tt VICE} adopts linear interpolation of the yields themselves.

\subsection{Absolute Yields} \label{subsec:abs_yields}

\begin{figure*}[!htb]
 \includegraphics[width=\textwidth]{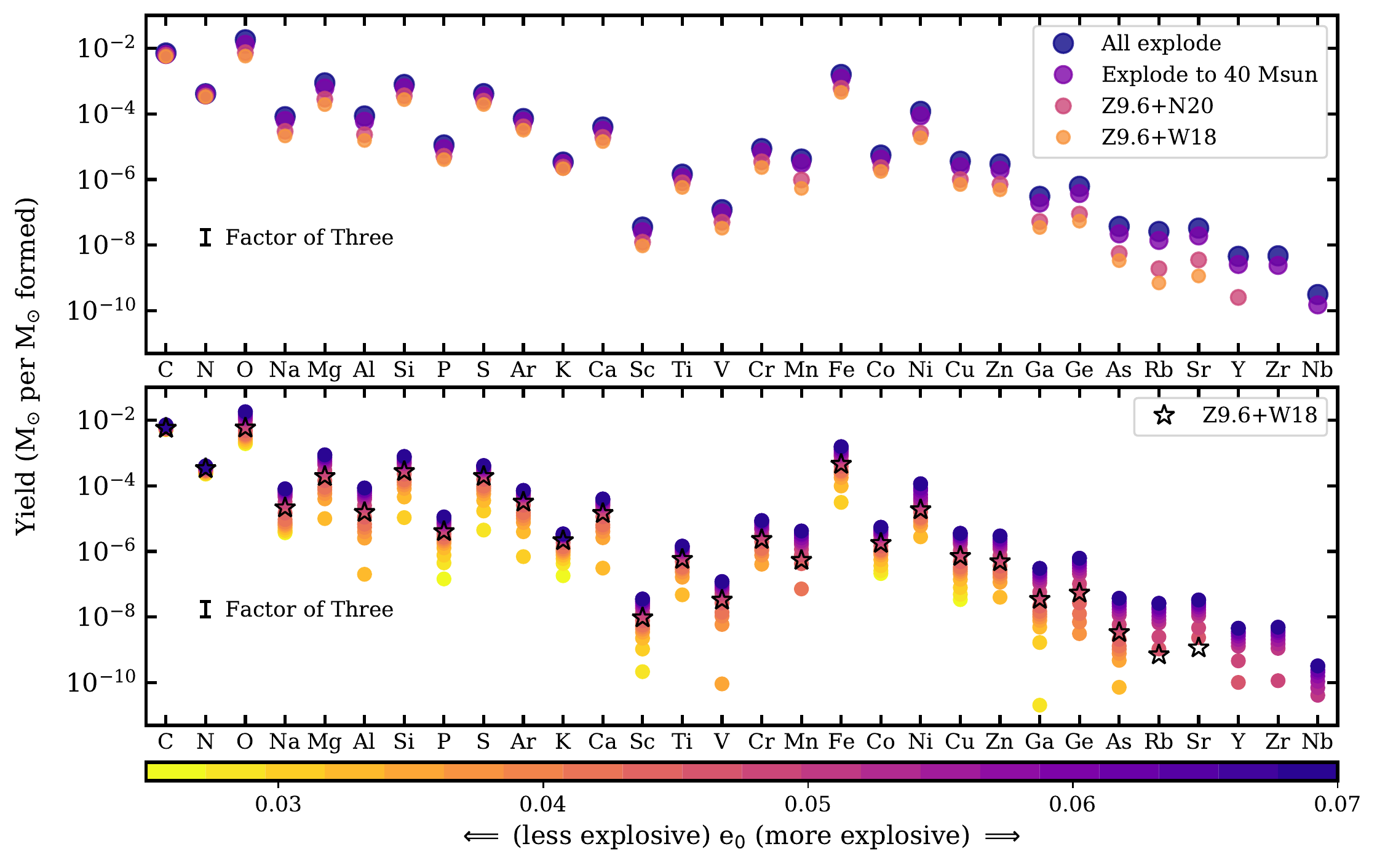}
 \caption{\textit{Top:} IMF-averaged yield for the All Explode model (dark blue), Explode to 40$\Msun$ (purple), Z9.6+N20 (pink), and Z9.6+W18 (orange) yield models for a range of elements. \textit{Bottom:} IMF-averaged yields for the range of $e_0$ values displayed in Figure~\ref{fig:e0_mass}. Darker/purple points indicate a larger $e_0$ (more explodability) and lighter/yellow points indicate a smaller $e_0$ (less explodability). The hollow black star shows the yields for the Z9.6+W18 set as a comparison (same as top panel). Both panels contain a bar denoting a factor of three for comparison. If an element does not have a model plotted, that model produces a negative net yield.}
 \label{fig:absYield}
\end{figure*}

We plot the IMF-averaged yields for 30 elements, spanning C to Nb, in Figure~\ref{fig:absYield}. The yields are in $\Msun$ of element produced per $\Msun$ of stars formed, making them dimensionless. The top panel includes the IMF-averaged yields for the Z9.6+W18, Z9.6+N20, Explode to 40$\Msun$, and All Explode landscapes. The absolute net yields of these models for all elements included in \citetalias{sukhbold2016} can be found in Table~\ref{tab:yield}. The bottom panel of Figure~\ref{fig:absYield} shows the range in IMF-averaged yields obtained by our 20 continuous explosion landscapes. 

Through this figure, we can compare two complex BH landscapes with a landscape with an upper mass explosion boundary and a landscape with everything exploding.  We see that the Z9.6+W18 and Z9.6+N20 landscapes produce similar absolute yields for most elements. The additional explosions in Z9.6+N20 only increase yields relative to Z9.6+W18 by a small amount. We see larger differences between the Z9.6+W18/N20 landscapes and the Explode to $40\Msun$ models, with variations exceeding a factor of 3 for many elements. These differences are largest for the weak $s$-process elements, but substantial for Na, Mg, Al, and the Fe-peak. The All Explode and Explode to 40$\Msun$ yields are strikingly similar, with no element showing variation larger than a factor of two. At least with the mass loss prescription adopted by \citetalias{sukhbold2016}, the explosive yields of nearly all elements decline for $M > 40\Msun$ (see Figure~\ref{fig:mass_yield_all}), so with a declining IMF these stars cannot contribute much to the integrated yield. While it is unlikely that all stars successfully explode, the All Explode model gives us an upper limit on the potential elemental yields produced by CCSN.

In the bottom panel of Figure~\ref{fig:absYield} we show the absolute yields calculated for 20 continuous explosion landscapes, as described in Sections~\ref{subsec:e0_preds} and~\ref{sec:IMF_int}. We plot the yields for all $e_0$ values and landscapes in Figure~\ref{fig:e0_mass} -- from very few explosions ($e_0=0.025$) to almost all stars exploding ($e_0=0.07$). The absolute yields span an order of magnitude or more for all elements except C and N, though much of this variation is for landscapes less explosive than Z9.6+W18. \add{In our subsequent analysis and discussion we will not consider landscapes less explosive than Z9.6+W18. These landscapes are too weak to explain the abundances of light $s$-process elements and fail to produce sufficient Si and O, (\citealp{brown2013}, \citetalias{sukhbold2016}). In support of these works, we find that the Z9.6+W18 with a Kroupa IMF has the minimum explodability necessary to reach solar O and Mg abundances (see Section~\ref{sec:discussion}).}

IMF-averaged yields shown here include winds that are released regardless of the ultimate fate of the core, but these only contribute significantly to He, C, N, O, F, Ne, and Na, four of which are shown in Figure~\ref{fig:absYield}. These elements have net wind yields that are comparable to or exceed the explosive yields for $\MZAMS \gtrsim 40\Msun$. While other elements have substantial wind yields at high masses, they are matched by the birth abundance. IMF-averaged net C and N wind yields exceed the IMF-averaged explosive yields by a factor of 3 and 1.5, respectively. The net O wind yields are $~10\%$ of the explosive O, and the net Na wind yields $\sim1\%$ of the explosive Na. Winds contribute less than $1\%$ to the net IMF-averaged yields of all other elements.

In both sets of absolute yields, all elemental abundances increase with increasingly explosive landscapes. C and N show very small changes between the explodability models. Their IMF-averaged yields in $\Msun$ per $\Msun$ formed are almost independent of explosion landscape or mass cutoff due to their dominant production in the stellar winds, independent of explodability. The constant absolute yield of C and N could make them very useful as observational constraints on the BH landscape, provided one can adequately isolate the CCSN contribution to their abundance. We find some variation between the Z9.6+W18 and Explode to 40$\Msun$ landscapes for the light-$Z$ elements, a larger amount for the $\alpha$ and Fe-peak elements, and the greatest variation in the weak s-process elements. We do not show the W18 and N20 yields for the heaviest elements as the birth abundances exceed the gross yields.

\subsection{The O/Mg Problem: Overproduction of O or Underproduction of Mg?}\label{subsec:OMg}

In this section we take a closer look at the IMF-averaged yields of O and Mg, two elements produced almost entirely by CCSN \citep{andrews2017, rybizki2017} that can be robustly measured by spectroscopic surveys. Their pure CCSN origin implies that solar metallicity CCSN models should reproduce the solar O to Mg ratio. However, this is not the case: solar metallicity CCSN models underproduce Mg relative to O. \citet{lc18} see Mg underproduction in all of their yield sets and find that it is worsened by the inclusion of stellar rotation. \add{This is further supported by \citet{prantzos2018}, who find significant Mg underproduction in their Galactic chemical evolution model that employs the \citet{lc18} yields.} Mg underproduction relative to O can also be expressed as O overproduction relative to Mg. In \citet{griffith19}, we find that the Z9.6+W18 yield set from \citetalias{sukhbold2016} and the yields used by \citet{rybizki2017} overproduce O relative to Mg by a factor of $1.5 - 2$. 

\begin{figure}[!htb]
 \includegraphics[width=\columnwidth]{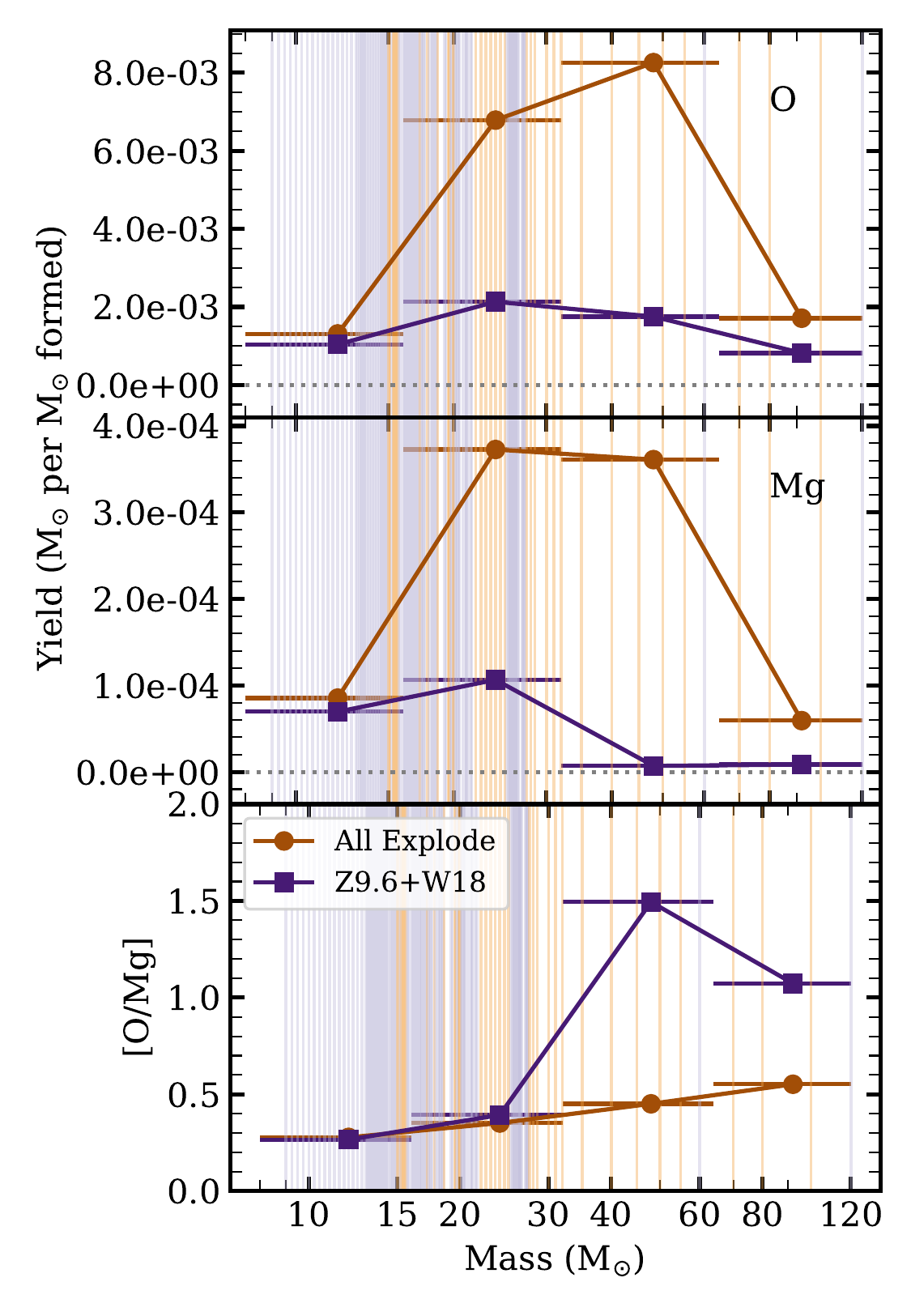}
 \caption{\add{Top and middle:} O (\add{top}) and Mg (\add{middle}) yields (in $\Msun$ per $\Msun$ formed) integrated under ranges of the stellar IMF. Points indicate the total yield from stars of mass $8-16\Msun$, $16-32\Msun$, $32-64\Msun$ and $64-120\Msun$ for the Z9.6+W18 explosion yield set (purple squares) and the All Explode yield set (orange circles).  Background colored lines indicate successful explosions under the Z9.6+W18 engine (light purple) and forced explosions (light orange). The grey dotted line denotes a yield of 0. \add{Bottom: [O/Mg] ratio calculated for each mass bin.}}
 \label{fig:omg_chunk}
\end{figure}

What is causing this discrepancy between O and Mg? To understand which mass ranges are contributing to the problem in our yields, we calculate the IMF-averaged abundances in four mass ranges: $8-16\Msun$, $16-32\Msun$, $32-64\Msun$ and $64-120\Msun$. We plot the the resulting net yields \add{in the top (O) and middle (Mg) panels of} Figure~\ref{fig:omg_chunk} for the Z9.6+W18 and All Explode cases. \add{The bottom panel shows the [O/Mg] ratios calculated for the yield in each mass bin, where
 \begin{equation}\label{eq:xy}
        [\text{X}/\text{Y}] = \log_{10}(\text{X}/\text{Y}) - \log_{10}(\text{X}/\text{Y})_{\odot}.
    \end{equation}
    }
    
We find that all four mass ranges significantly contribute to the total O abundance for both landscapes, with stars of  $16-32\Msun$ and $32-64\Msun$ dominating the production. Though few stars in the two largest mass bins explode under the Z9.6+W18 model, the O yields are non-zero due to significant wind yields from high mass stars. Mg, however, has insignificant net wind yields, causing IMF-averaged abundances of the W18 landscape to be negligible for the stellar mass ranges $32-64\Msun$ and $64-120\Msun$. The Z9.6+W18 yields slightly overproduce O relative to Mg for stars of $8-16\Msun$ and $16-32\Msun$, but they drastically overproduce O in the $32-64\Msun$ and $64-120\Msun$ mass bins.

When all stars explode, explosive nucleosynthesis from high mass stars raises both the O and Mg yields. The impact is larger for Mg because O wind yields would be present even without explosions. Thus, while O is still overproduced, the severity of the overproduction is lessened. We find O overproduction (or Mg underproduction) by a factor of 3.7 for the Z9.6+W18 landscape and a factor of 2.5 for the All Explode scenario, assuming the \citet{asplund2009} solar abundances discussed below (see Table~\ref{tab:solar}). While at face value this argues for a landscape with minimal BH production, the likelihood of all stars exploding is slim. We suspect that resolving the O/Mg problem will require a change to the nucleosynthesis predictions in the pre-SN phase of massive star evolution, but the origin of this physics is not obvious. We return to this issue in Section~\ref{subsec:disc_disrepancies} below, where we conjecture that the resolution will be one that both decreases the O yield and increases the Mg yield.

The dependence of the IMF-averaged O/Mg ratio on explodability could provide a mechanism for metallicity dependent [O/Mg] trends. While IR studies, such as APOGEE, find no metallicity trend in [O/Mg] ratios \citep{weinberg2019}, optical studies, like GALAH, find that [O/Mg] decreases as a function of increasing [Mg/H] (\citealp{griffith19}; see also, e.g., \citealp{bensby2014}). This disagreement on the metallicity dependence of O abundances may stem from 3D non-LTE effects afflicting optical abundances derived from the O triplet (OI 7772, OI 7774, and OI 7775\AA) \citep[e.g.][]{kiselman1993, amarsi2015} or from systematics in modeling the molecular effects on the OH and CO lines, used to derive O abundances in the IR \citep[e.g.][]{collet2007, hayek2011}. Observationally, the existence of a metallicty dependent [O/Mg] trend remains uncertain. 

At fixed mass, the \citet{chieffi2004} and \citet{limongi2016} models predict only modest dependence of O and Mg yields on metallicity \citep[see Figure 19 of][]{andrews2017}. However, if BH formation is more prevalent at lower metallicity as some theoretical arguments suggest \citep{pejcha2015, raithel2018}, then the changing explodability landscape would cause higher O/Mg ratios at lower [Mg/H]. It is difficult to say whether this trend could be large enough to explain the trend found in optical abundance studies without knowing the resolution to the overall O/Mg normalization conundrum. \add{The \citet{chieffi2004} and \citet{limongi2016} yields also show a mild metallicity dependence for Si and Ca as well as a stronger metallicity dependence for Al, Na, and Co in the IMF-averaged yields from \citet{andrews2017}. Many of these elements are also observed to have metallicity dependent trends in APOGEE and GALAH \citep[e.g.,][]{weinberg2019, griffith19}. We look forward to investigating the metallicity dependence of CCSN yields from complex BH landscapes with neutrino powered explosions as they become available.} 

\subsection{The O/Mg Solar Abundance Conundrum} \label{subsec:solar}

Our work in the subsequent sections takes the recommended photospheric solar O value from \citet{asplund2009}, $\log{\epsilon_{\text{O}}} = 8.69$\footnote{$\log{\epsilon_{\text{X}}} = \log(N_{\text{X}}/N_{\text{H}}) + 12$, where $N_{X}$ is the number density of element $\text{X}$.}.
However, as the authors note in their review, the solar O value has been disputed for the last three decades. Today a $~0.15$ dex range still exists between works of different analysis methods. For example, asteroseismic O values from \citet{villante2014} find a higher $\log{\epsilon_{\text{O}}} = 8.85$. While the O abundance is disputed, the solar Mg value is more settled upon in literature, with variations of only $\sim 0.05$ dex. 

The uncertainty in the solar O value leads to some uncertainty in the [O/Mg] value.
In Table 1 we list the O and Mg values from \citet{grevesse1998}, \citet{lodders2003}, \citet{asplund2009}, and \citet{lodders2010} and calculate the [O/Mg] value for the Z9.6+W18 yields with each set of solar abundances. \citet{caffau2008} recommend an O abundance of $\log{\epsilon_{\text{O}}} = 8.76$, but they do not calculate the solar Mg value so we do not include this source in the table.

\begin{table}
\centering
\caption{Table of solar O and Mg values ($\log{\epsilon_{\text{X}}}$) from \citet[][GS98]{grevesse1998}, \citet[][L03]{lodders2003}, \citet[][AGS09]{asplund2009}, and \citet[][L10]{lodders2010}. [O/Mg] abundances are calculated with W18 yields according to Equation~\ref{eq:xy} and included in the final column. \label{tab:solar}}
\begin{tabular}{lrrr}
\toprule
   Set &     O &    Mg &  [O/Mg] \\
\hline
  GS98 &  8.83 &  7.58 &   0.403 \\
   L03 &  8.69 &  7.55 &   0.513 \\
 AGS09 &  8.69 &  7.60 &   0.563 \\
   L10 &  8.73 &  7.54 &   0.463 \\
\hline
\end{tabular}
\end{table}

We find a range of predicted [O/Mg] from 0.40 to 0.56. The low solar O value from \citet{asplund2009}, assumed here, returns the highest [O/Mg] value and thus the largest implied O overproduction. Taking a higher solar O abundance such as \citet{grevesse1998} could alleviate $0.1-0.15$ dex of this discrepancy, but would not completely resolve it.  
We continue to use solar values from \citet{asplund2009} in our subsequent work, but we recognize that a change in the solar abundance set may lessen the [O/Mg] discrepancy.

\section{Abundance Ratio Comparison to Data}\label{sec:ab_ratios}

\begin{figure*}[!htb]
 \includegraphics[width=\textwidth]{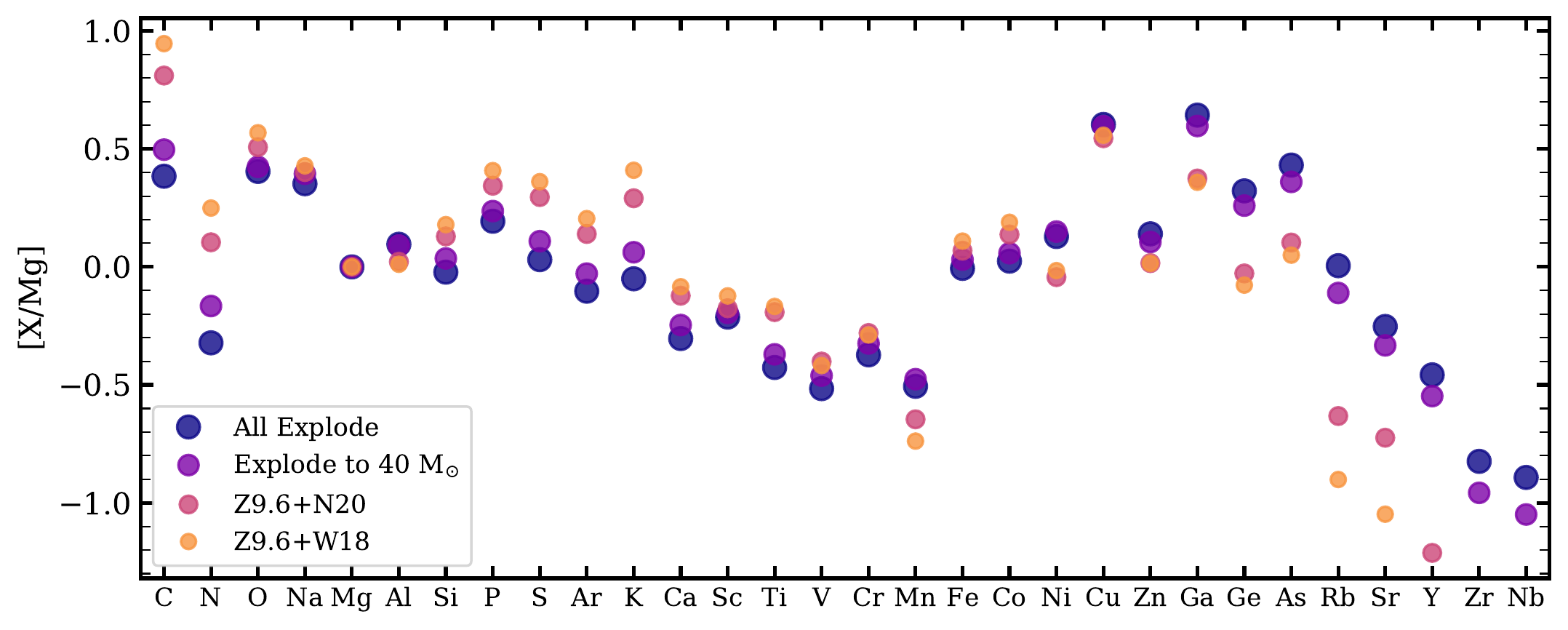}
 \caption{[X/Mg] relative abundances for the All Explode (dark blue), Explode to 40$\Msun$ (purple), Z9.6+N20 (pink), and Z9.6+W18 (orange) BH landscapes. [X/Mg] values are calculated with solar values from \citet{asplund2009}. If an element does not have a model plotted, that model produces a negative net yield.}
 \label{fig:xmg}
\end{figure*}

Using the absolute yields in Table~\ref{tab:yield}, we calculate abundance ratios for a variety of BH landscapes. We plot the [X/Mg] values for the Z9.6+W18, Z9.6+N20, Explode to 40$\Msun$, and All Explode landscape in Figure~\ref{fig:xmg}. In this section we will only show and discuss the [X/Mg] and [X/O] values, but other ratios can be calculated using Equation~\ref{eq:xy} and the absolute yields (Table~\ref{tab:yield}). We choose to use Mg as our primary reference because it is a pure CCSN element \citep{andrews2017} that can be observed with accuracy and precision. It is used as a reference element to empirically derive the CCSN and SNIa contribution to elements in APOGEE \citep{weinberg2019} and GALAH \citep{griffith19}. We would expect a successful set of CCSN yields to achieve $\xmg$ abundances near solar for $\alpha$-elements, whose production is dominated by CCSN, and slightly sub-solar for those with more SNIa production, such as the Fe-peak elements. 

The absolute Mg yields (Figure~\ref{fig:absYield}) are dependent upon the BH landscape. They change by a factor of $\sim4.5$ between the Z9.6+W18 and All Explode models. Therefore, the [X/Mg] value for an element whose absolute yields also depend on the BH landscapes will reflect a combination of Mg and X's landscape dependence. If the yields of element X have the same BH landscape dependence as those of Mg, then the [X/Mg] values should not change with the changing BH landscape. This is the case for Ne, Na, Al, and Sc, which show only $\sim0.1$ dex differences between the Z9.6+W18 and All Explode landscapes. 

If an element shows less variation in absolute yield than Mg, then the [X/Mg] values decrease with increasing explodability (blue points are lowest in Figure~\ref{fig:xmg}). We see this behavior for C, N, O, F, Si, P, S, Cl, Ar, K, Ca, Ti, V, Cr, Fe, and Co. The smaller the dependence of an element's absolute yields on explodability, the larger the range in [X/Mg] the models span. The light-$Z$ elements have variations of $\sim 0.3-0.6$ dex between the Z9.6+W18 and All Explode landscapes while the Fe-peak elements span smaller ranges of $\sim 0.2-0.3$ dex. C and N, which show almost no variation in absolute yield with explosion landscape, display the largest difference in [X/Mg] amongst this group, with $\sim0.7$ dex between the [X/Mg] values of the Z9.6+W18 and All Explode landscapes.
    
Alternatively, if an element shows more variation in absolute yield than Mg, then the [X/Mg] value increases with increasing explodability (blue points are highest in Figure~\ref{fig:xmg}). This behavior is seen for Mn, Ni, Ce, Zn, Ga, Ge, As, Se, Br, Kr, Rb, Sr, Y, Zr, and Nb. The Fe-peak and Fe-cliff\footnote{We define the ``Fe-cliff'' as elements on the steeply dropping edge just above the Fe-peak.} elements show smaller variations, with $\sim0.2-0.4$ dex between the least and most explosive landscape. The weak $s$-process elements span up to 1 dex, the largest range of any element studied here.

The best elemental ratio to constrain the BH landscape will be that of two elements that have different absolute yield dependences on explodability. The two elements' yields must be able to be modeled accurately and observed with accuracy and precision. C and N may be interesting reference elements to use in such a comparison as their absolute abundances show very little dependence on BH landscape, as was briefly mentioned in Section~\ref{subsec:abs_yields}. The [X/C] or [X/N] values for an element, X, that \textit{does} vary with BH landscape could help us place observational constraints on the true explodability landscape in the MW. C and N, however, have other sources of production, such as in AGB stars \citep{andrews2017}. Their use in constraining the CCSN BH landscape requires the ability to separate the CCSN component from the delayed component in abundance measurements. \citet{weinberg2019} and \citet{griffith19} empirically separate the CCSN and SNIa components of APOGEE and GALAH elements, respectively, with the latter including C. However, this model is calibrated to delayed enrichment from SNIa, not AGB stars. If we are able to convincingly isolate the CCSN contribution to C and N, they could be valuable diagnostics of BH formation, with the complication (see Section~\ref{subsec:disc_disrepancies}) that their yields could change if high mass stars experience less mass loss and collapse to BHs \textit{without} driving enriched winds. 
    
In this paper we only employ Mg and O as reference elements in our comparison to observational data. \add{Rather than compare directly to solar abundances, which represent a mixture of CCSN and SNIa contributions, we use results from \citet{griffith19} and \citet{weinberg2019} to infer the fraction of each element at solar abundances that arises from CCSN. Theese papers determine median abundance trends for high-[Mg/Fe] and low-[Mg/Fe] disk populations measured by GALAH DR2 \citep{griffith19} and APOGEE DR14 \citep{weinberg2019}, then fit these trends with an empirical two-process model that describes the abundances as the sum of a CCSN and SNIa component. Roughly speaking, the [Mg/Fe] value is used to infer the fraction of Fe coming from SNIa on the two median sequences, and the separation of the two sequences in [X/Mg] is used to infer how much of element X comes from SNIa. We primarily use results from GALAH because it has more elements available, but we use the APOGEE results for S as it is not included in GALAH DR2.}

\add{We discuss the two-process model in more depth in Appendix~\ref{ap:2proc}. Although there are uncertainties from both the abundance measurements and assumptions in the two-process model, isolating the CCSN component of abundances in this way is an improvement over comparing CCSN models to uncorrected abundances. We take the solar abundances themselves from \citet{asplund2009}; the role of the two-process model is just to determine the fraction of each element that we attribute to CCSN.}

\begin{figure*}[!htb]
 \includegraphics[width=\textwidth]{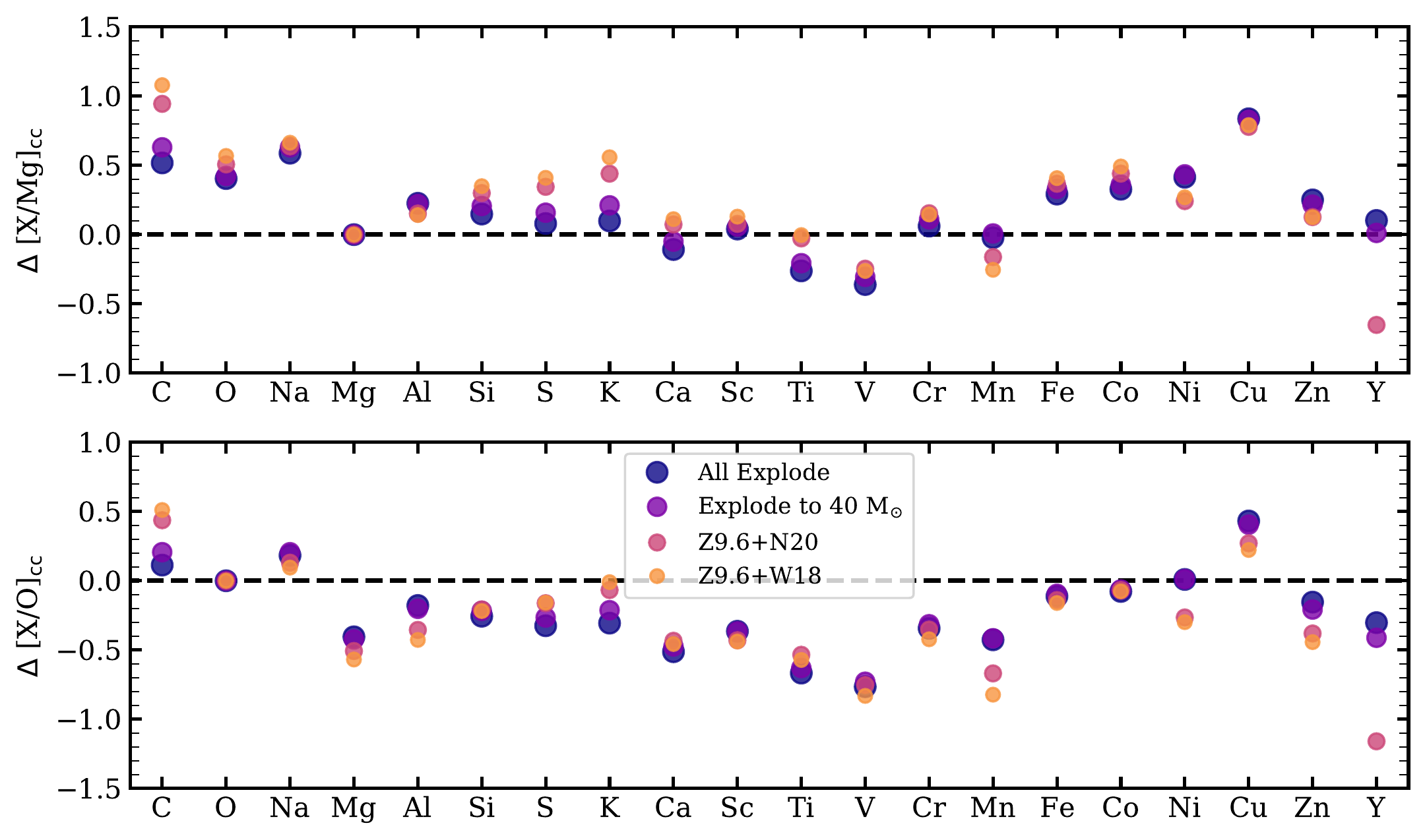}
 \caption{Offset by which each explodability landscape over/underpredicts the CCSN contribution to [X/Mg] (top) and [X/O] (bottom, with Z9.6+W18 in yellow, Z9.6+N20 in orange, Explode to $40\Msun$ in purple, and All Explode in dark blue. The dashed line at 0.0 represents the point where theoretical and empirical results agree. Observational constrains are taken from \citet{griffith19} for all elements but S, which is taken from \citet{weinberg2019}. The top panel is more relevant if the O/Mg problem (Section~\ref{subsec:OMg}) arises because models overproduce O, and the lower panel is more relevant if the models are accurate for O but underproduce Mg.}
 \label{fig:fracCCSN}
\end{figure*}

Figure~\ref{fig:fracCCSN} shows the ratio of the BH landscapes' CCSN yields of element X to that of Mg (top) and O (bottom), divided by the ratio of estimated CCSN yield to that of Mg from \add{the two-process model} \citep{griffith19}. \add{Our $\Delta [\text{X}/\text{Mg}]_{\text{cc}}$ and $\Delta [\text{X}/\text{O}]_{\text{cc}}$ represent} the difference between the theoretical and empirical CCSN [X/Mg] and [X/O] values. \add{We define $\Delta [\text{X}/\text{Mg}]_{\text{cc}}$ and $\Delta [\text{X}/\text{O}]_{\text{cc}}$ in terms of the theoretical yield and two-process parameters in Appendix~\ref{ap:2proc}.}

If a BH landscape model has a $\Delta \xmg_{\text{cc}} = 0$, then it reproduces the GALAH CCSN yields relative to Mg from \citet{griffith19}. If the value is above/below 0, then the model over/underpredicts the inferred CCSN yield by the indicated number of dex. The Z9.6+W18 points reproduce the results shown in Figure 17 of \citet{griffith19}. As in \citet{griffith19}, we see substantial overprediction of C, O, Na, K, and Cu yields, relative to Mg, for the Z9.6+W18 landscape. While these discrepancies lessen for more explosive models, they do not disappear. \citetalias{sukhbold2016} show super-solar [C/O] values for the Z9.6+W18 and Z9.6+N20 models. The offset worsens in our comparison, as we attribute $25\%$ of C production to delayed sources \citep{griffith19}. Similarly \citetalias{sukhbold2016} find super-solar Na, Co, and Cu abundances for the Z9.6+W18 yields. Their discrepancies are larger here as all three elements have substantial delayed components. We include Y as an example of a neutron capture element. Our net Y yields are substantially underpredicted for the Z9.6+N20 landscape, in agreement with \citetalias{sukhbold2016}'s comparison to solar. 
The quantitative degree of underproduction is uncertain because the two-process decomposition implicitly assumes that the delay-time distribution of non-CCSN Y production follows that of SNIa, where the actual source of delayed Y is likely AGB stars.

Overall, we find that more explosive models agree better with observational results, as the disparity seen in C, O, and K shrinks for the All Explode model. This phenomenon is more thoroughly explained for O in Section~\ref{subsec:OMg}, though we note that no explosion landscape can completely correct for the O overproduction. Furthermore, no BH landscape alleviates the tension between the theoretical and observational results for Na or Cu.

The O/Mg discrepancy described in Section~\ref{subsec:OMg} and seen in Figure~\ref{fig:fracCCSN} could be caused by overproduction of O or underproduction of Mg (or a combination of the two). Removing the effect of this discrepancy from the top panel of Figure~\ref{fig:delta_xmg} does not translate to a constant additive or multiplicative offset, so we plot $\Delta [\text{X}/\text{O}]_{\text{cc}}$ in the bottom panel for completeness. The observational CCSN fractions are again inferred from GALAH Mg, but we take O yields to be the theoretical reference. Here we see that the elements whose abundances were overpredicted relative to Mg (e.g. Na, K, Cu) better reproduce the empirical results when normalized to O. We still see C and Cu overproduction when scaling to O as these elements have a larger offset than O in the top panel. However, assuming that the O/Mg resolution lies in Mg underproduction also leads to underprediction of Mg, Ca, Sc, Ti, V, Cr, and Mn. The Fe-peak elements come into better agreement with the observed abundances if we assume Mg underproduction (lower panel) rather than O overproduction (upper panel). As a simplified characterization of overall agreement, we note that the median absolute value of $\Delta \xmg_{\text{cc}}$ and $\Delta[\text{X}/\text{O}]_{\text{cc}}$ for the elements plotted in Figure~\ref{fig:fracCCSN}, excluding Y, are 0.27 and 0.42, respectively, for the Z9.6+W18 model, dropping 0.25 and 0.31 for the All Explode model. If we compute the mean absolute deviation instead of the median, we find 0.36 and 0.35 for the Z9.6+W18 model and 0.27 and 0.30 for the All Explode model. Thus, a characteristic level of disagreement with observed abundance ratios is $\sim 0.3-0.4$ dex, and moderately worse for the Z9.6+W18 model than for the All Explode model.

\subsection{Upper Mass Cut vs. Black Hole Landscape}

Neutrino powered explosions \citepalias[e.g.][]{sukhbold2016} produce BH landscapes with regions of explodability and regions of BH formation. This model is distinct from treatments that adopt a cutoff $\MZAMS$ above which all stars collapse to BHs \citep[e.g.][]{lc18}. In this section we explore the abundance signatures produced by landscapes with islands of explodability and those with upper mass cutoffs in search of elements that could observationally distinguish between the two scenarios.

Figure~\ref{fig:fracCCSN} compared the abundance ratios from landscapes with an upper mass cut (Explode to 40$\Msun$) to those those with islands of explodability (Z9.6+W18 and Z9.6+N20). This is not the best comparison, as the overall difference in the scale of explodability varies between them, with the Explode to 40$\Msun$ landscape producing higher yields across all elements. We can level the playing field by comparing landscapes with an upper mass cut to those with islands of explodability that produce the same amount of Mg and O. We then analyze how the suite of all abundance ratios differs between the two landscapes.

To compare the Z9.6+W18 landscape with an upper mass cutoff landscape, we first find the explodability mass cuts that produce the same yields of Mg and O as Z9.6+W18. We iteratively integrate explodability functions of increasing upper mass cut with {\tt VICE} until we converge on a landscape that is within 1\% of the Z9.6+W18 yield. We find that a landscape with successful explosions for stars with $\MZAMS < 21.9\Msun$ best reproduces the Z9.6+W18 O yield of $5.75\times 10^{-3} \, \Msun$ per $\Msun$ formed, and $\MZAMS <21.0\Msun$ best reproduces the Z9.6+W18 Mg yield of $1.91\times 10^{-4} \, \Msun$ per $\Msun$ formed. Because the masses are very similar, we only show and discuss the 21$\Msun$ upper limit here. The Explode to 21$\Msun$ landscape differs from Z9.6+W18 through its inclusion of explosive yields for stars near $\MZAMS$ of 15 and 20$\Msun$, which collapse in Z9.6+W18, and through its exclusion of explosive yields from stars with $\MZAMS$ of 25, 60, and 120$\Msun$, which explode in Z9.6+W18. 

\begin{figure*}[!htb]
 \includegraphics[width=\textwidth]{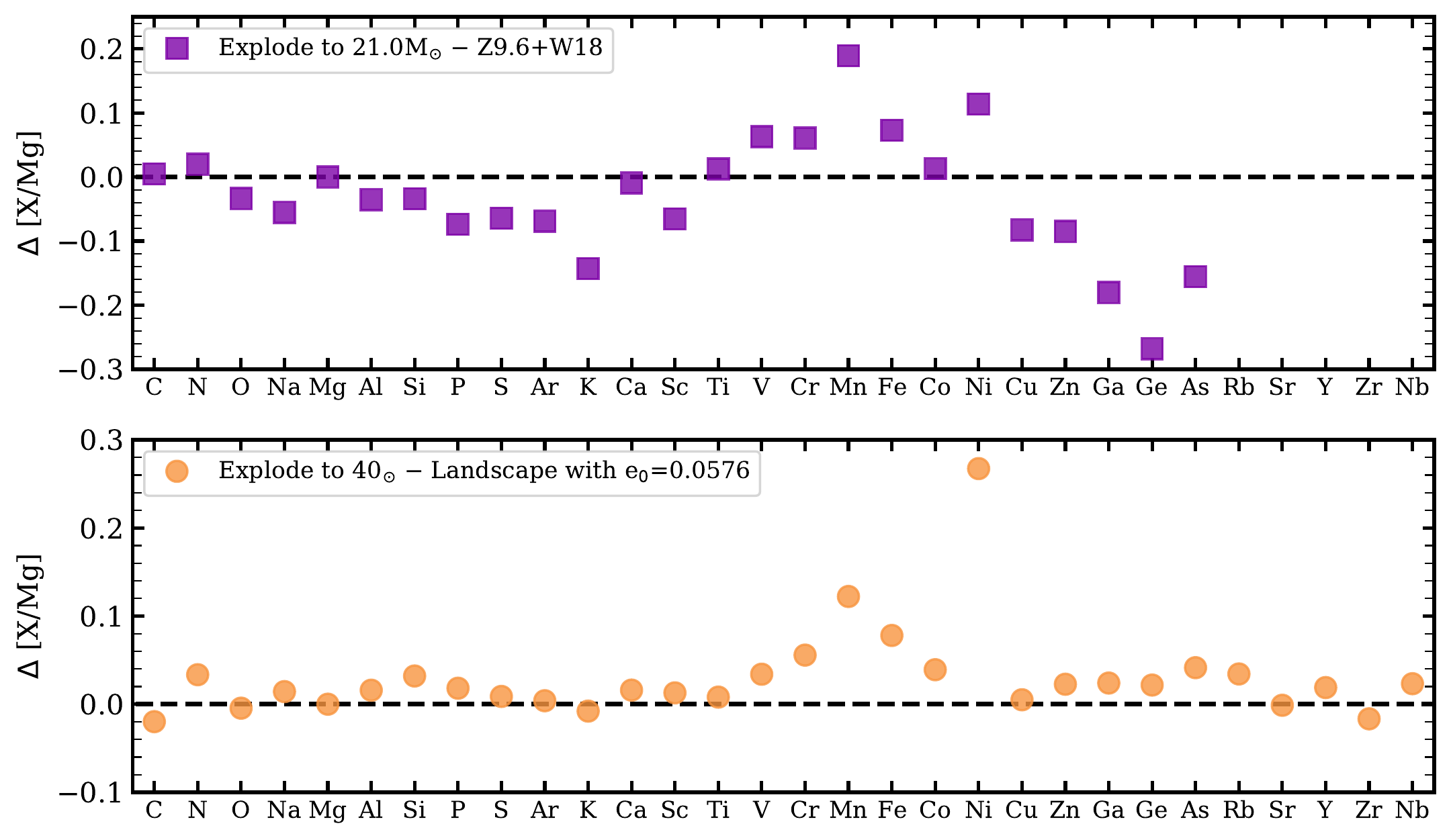}
 \caption{Differences between IMF-averaged yield ratios for landscapes with an explodability upper mass vs. those with regions of explosion and implosion. \textit{Top:} Comparison of yields from the Z9.6+W18 and a landscape with an upper mass cut that reproduces the Z9.6+W18 Mg yield (21$\Msun$). \textit{Bottom:} Comparison of yields from the Explode to 40$\Msun$ and an $e_0$ landscape that reproduces the Explode to 40$\Msun$ Mg yield ($e_0 = 0.0576$).}
 \label{fig:delta_xmg}
\end{figure*}
    
We calculate the yields and abundance ratios for the Explode to 21$\Msun$ landscape as described above. We plot the difference in the [X/Mg] ratio between it and the Z9.6+W18 landscapes in the top panel of Figure~\ref{fig:delta_xmg}. No values are plotted for the five heaviest elements as one or both of the yields return a negative value. We find very small differences ($<0.1$ dex) for most of the $\alpha$ and light-$Z$ elements, whose explosive yields are dominated by $8-20\Msun$ stars. Larger differences occur for K, Mn, Ni, and the weak $s$-process elements. We find that the Z9.6+W18 yields produce a [K/Mg] ratio 0.15 dex higher than the Explode to 21$\Msun$ yields, likely due to substantial contribution from stars around 25$\Msun$. The Explode to 21$\Msun$ landscape shows higher Mn and Ni, with $\Delta \xmg \approx 0.1-0.2$ dex. Both elements have high explosive yields near 15$\Msun$, a mass range that collapses to BHs in the Z9.6+W18 model. Finally, we find higher abundances of the Fe-cliff and weak $s$-process elements from the Z9.6+W18 landscape, which includes important contributions from higher mass stars. 

For a more explosive comparison, we plot the difference between the Explode to 40$\Msun$ abundances (shown in Figure~\ref{fig:xmg}) with those from an $e_0$ explosion landscape (Section~\ref{subsec:e0_preds}) with the same Mg yields. To produce such a landscape, we interpolate between the yields of 100 landscapes with $e_0$ of $0.25-0.07$. We find that a landscape with $e_0 = 0.0576$ best reproduces the Mg yield of the Explode to 40$\Msun$ model. An $e_0 = 0.0573$ best reproduces the O yield, but we again omit this case due to similarity. As seen in Figure~\ref{fig:e0_mass}, such a landscape produces explosions at most masses, with small regions of BH formation at $\MZAMS$ of 15, 22, 27, and $35-50 \Msun$. We find that the abundance ratios produced by the Explode to 40$\Msun$ and $e_0=0.0576$ landscapes are very similar, with most elements showing abundance differences $< 0.05$ dex. The largest differences again appear for Mn and Ni. These two elements have high production near 25$\Msun$ and $35-40\Msun$ -- regions of high compactness parameters that do not explode in the $e_0$ model. 

In both cases, the elements that distinguish landscapes with islands of explodability from those with upper mass cuts are those whose yields have sharp peaks for mass ranges that collapse in one scenario but not in the other, often associated with high compactness parameter. While most abundances show little difference between these two scenarios, several show differences of $0.05-0.3$ dex. At present, it is difficult to say whether a complex landscape is empirically favored by observed abundance ratios or not, in part because none of our models produce good across-the-board agreement with the data (Figure~\ref{fig:fracCCSN}). In addition, empirical abundance scales often have systematic uncertainties at the $0.05-0.1$ dex level, and the isolation of the CCSN contribution to abundances is uncertain. Nonetheless, useful tests appear within reach of improving theoretical models and observational surveys over the next several years. 

\section{Discussion}
\label{sec:discussion}

It is evident from Figure~\ref{fig:fracCCSN} that none of the models examined here achieves a good across-the-board fit to all of the observationally inferred yields.  In Section~\ref{subsec:disc_theory} we discuss some of the sources of uncertainty in the theoretical models, then turn in Section~\ref{subsec:disc_disrepancies} to potential origins of some of the larger discrepancies with observations.  The latter discussion is necessarily speculative, as we do not have detailed calculations for alternative yield models.  We discuss the impact of the stellar IMF in Section~\ref{subsec:disk_imf}, where we show that plausible changes to the IMF can have a large impact on the absolute yields but only limited impact on predicted abundance ratios.

\subsection{Theoretical Uncertainties}
\label{subsec:disc_theory}

One important source of uncertainty in both explodability and nucleosynthesis calculations is the mass loss prescription for progenitor stars. In Figure~\ref{fig:mass_yield} and in Figure~\ref{fig:mass_yield_all} below, the yields of most elements show a clear transition at $40\Msun$.  As discussed in Section~\ref{subsec:yield_v_mass}, this transition arises because the pre-SN models from S16 with $M>45\Msun$ have shed their entire H shell and are starting to shed their He and C shells. These massive stars release He, C, N, and O in winds rather than continuing fusion to heavier elements. A more severe mass loss prescription would shift this transition down to lower $\MZAMS$, but the recipe used in our models is probably at the high end of what is observationally and theoretically allowed \citep[e.g.,][]{yoon2017,vink2017,sander2020}.  

Conversely, less severe mass loss would shift the peaks of the explosive yields of He, C, N, and O to higher masses. The maximum O yield would occur around 60 or 70$\Msun$ with less decline at higher masses. Yields of Fe-peak elements are not sensitive to mass loss and thus would not change. However, a reduction of mass loss would increase the core mass of the most massive stars and thereby reduce their explodability -- affecting the IMF-averaged yields of all elements. The landscapes of \citetalias{sukhbold2016} find islands of explodability for the most massive stars (e.g. 60 and 120$\Msun$ in Z9.6+W18). With weaker mass loss at $\MZAMS > 45 \Msun$, these islands might shrink or disappear entirely.

As emphasized by, e.g., \citet{Patton2020} and \citet{Laplace2021}, binary star evolution can strongly affect the core structure of massive stars by accelerating the stripping of their outer layers. Changes to the core profile and to the pressure on outlying fusion shells can have an important impact on explodability and nucleosynthesis. Comprehensive models that account for these effects are not yet available. Possible effects relevant to our results include the stripping of the H envelope and reduction of the binary stars' core mass. The changes in core structure will shift the peak in compactness parameter vs. $\MZAMS$ to a higher mass ($\sim 27 \Msun$) and thus increase the explodability of lower mass stars.

The evolved structure of pre-SN cores can change sharply with small changes of initial conditions because of mergers and mixing between fusion shells, as emphasized by \cite{sukhbold2018}. This sensitivity is the underlying reason that the predicted explosion landscape is complex. While it necessarily makes the predictions for any specific progenitor mass uncertain, this sensitivity may not change population-averaged predictions much, just altering the precise mapping between progenitor mass and explodability/yield.  However some changes, e.g., from stellar rotation or different convective overshoot prescriptions, could alter the predicted nucleosynthesis systematically by enhancing mixing between radial zones.  This mixing could enable fusion reactions that do not occur in current models because the necessary nuclei are not present at the same location when conditions that would enable fusion arise. Discrepancies between predicted and empirically inferred yields could provide clues to this missing physics.

\subsection{Interpretation of Discrepancies}
\label{subsec:disc_disrepancies}

As discussed in Section~\ref{subsec:OMg}, the clearest discrepancy between our models and data is the relative yield of O and Mg.  A similar tension is present in the two yield models considered by \cite{rybizki2017}, though there it is at the level of $\sim 0.15$ dex (see their Figure 14) vs. $0.4-0.6$ dex in our models.  The production of both elements is expected to be dominated by CCSN and massive star winds, so uncertainties in the contribution of other processes to solar abundances are minimal. The discrepancy is present for all of our landscape scenarios, becoming more severe for landscapes with more BH formation, so simply changing which stars explode will not resolve the conflict in the relative abundances. 

A plausible solution might lie in the nuclear reaction rates of both the triple-$\alpha$ reaction that produces C from He and the $^{12}$C($\alpha$,$\gamma$) reaction that produces O. The relative value of these rates determines the C/O ratio during He burning. Either raising the triple-$\alpha$ rate or lowering the $^{12}$C($\alpha$,$\gamma$) rate would decrease O production, and the larger abundance of C would provide more $^{12}$C seeds for the production of $^{24}$Mg. Thus this solution would both reduce the O yield and raise the Mg yield at a given $\MZAMS$. Because these changes would alter the core structure, they would also affect explodability, so the full impact of a reaction rate change is difficult to assess without extensive calculations. \add{We note, however, that the works of \citet{farmer2019, farmer2020} and \citet{woosley2021} find that decreasing the $^{12}$C($\alpha$,$\gamma$) and increasing the triple-$\alpha$ reaction rate allows for the production of higher mass BHs before pair-instability supernovae set it. Their adjustments to the reaction rates produce a mass gap that is in better agreement with the LIGO results.}

Full exploration of this idea is beyond the scope of this paper, but in a preliminary investigation we have computed models for a selection of masses between 12 and $30\Msun$ in which we modify the He-burning reaction rates. The adopted rates in \citetalias{sukhbold2016} were from \citet{CF88} for triple-$\alpha$, and 1.2 times that of \citet{Buc96} for $^{12}$C($\alpha$,$\gamma$). The triple-$\alpha$ reaction rate was increased to the upper end of allowed values (\citealp{Kib20}, $\sim35$\% increase from \citealp{CF88}) while the $^{12}{\rm C}(\alpha,\gamma)$ reaction rate was decreased (\citealp{deBoer17}, $\sim15$\% decrease from \citealp{Buc96}), thus making stars with substantially more C rich cores at the time of C-ignition. In the All Explode scenario, this change mildly decreases O yields and significantly increases Mg yields such that the median $^{16}$O/$^{24}$Mg ratio in this mass range shrinks by a factor of $\sim 2.7$. Changes in the opposite direction, by keeping the triple-$\alpha$ rate the same and increasing $^{12}{\rm C}(\alpha,\gamma)^{16}$ by $\sim$40\%  have a much smaller net impact on the O/Mg ratio.

Regardless of whether we normalize O yields or Mg yields to solar abundance, C is overproduced by a substantial factor in the Z9.6+N20 and Z9.6+W18 models, and for all models with Mg normalization (Figure~\ref{fig:fracCCSN}. This overproduction is largely a consequence of the high C wind yields from massive stars. The conflict is larger here than in Figure 24 of S16 because we attribute 25\% of C to non-CCSN based on the GALAH two-process decomposition \citep{griffith19}. Although we do not show N in Figure~\ref{fig:fracCCSN} because we do not have a two-process decomposition, Figure~\ref{fig:xmg} suggests that it would be similarly overproduced for the same reason. The most obvious way to lower C and N is to reduce the winds from $\MZAMS > 40\Msun$ stars \textit{and} assume that reduced mass loss causes these massive stars to form BHs, so that the C and N are not released in supernovae explosions. As discussed in Section~\ref{subsec:disc_theory}, a weaker mass loss prescription is empirically plausible. A change in this direction would also reduce the yield of He and slightly reduce the yield of O (see discussion of wind yields in Section~\ref{subsec:abs_yields}.)

With the O normalized to solar, the ``heavy'' $\alpha$-elements Ca and Ti are underproduced to about the same degree as Mg.  Perhaps there is a common physical origin of these three discrepancies, though the intermediate $\alpha$-element Si is only mildly underproduced.  The odd-$Z$, Fe-peak elements V and Mn are also severely underproduced (less so for Mg normalization), even though we have attempted to remove the SNIa contribution to the solar abundances via the two-process decomposition. The production of Fe-peak elements depends on the treatment of stellar explosions in addition to the pre-SN evolution. Compared to 3D CCSN models \citep[see review by][]{janka2016}, calibrated 1D models of \citetalias{sukhbold2016} have overall shallower core bounce in their explosions. An earlier and deeper core bounce would result in a lower electron fraction $(Y_e)$ and a higher number of neutrons produced. A more neutron rich environment could support the increased production of odd-$Z$ and Fe-peak elements such as V and Mn. 

Cu is mildly overproduced for the O-normalized, Z9.6+N20 and Z9.6+W18 landscape models, but the discrepancy is substantial for the All Explode or Explode to 40$\Msun$ models, or for any of the models with Mg normalization. Cu is dominantly produced by the $s$-process in the He burning shell. Cu production could be altered by changing the reaction rate for the neutron source, $^{22}$Ne($\alpha$,n), which is currently uncertain. Such a change would also decrease production of Zn, Y, and some Fe-peak isotopes, so this solution could be corroborated if the CCSN contribution to these elements/isotopes could be adequately isolated. Cu overproduction might also be rectified by changing the mass loss prescription, since Cu is produced in the He shell. 

The underproduction of solar Sr, Y, and Zr (Figure~\ref{fig:xmg}) is almost certainly a consequence of other processes contributing to these elements. The two-process modeling of GALAH abundances \citep{griffith19} and abundances of these elements in low metallicity stars \citep{zhao2016, mishenina2019, chaplin2020} imply that there is a substantial ``prompt'' contribution to these elements in addition to $s$-process production in AGB stars.  Based on the calculations of \cite{Vlasov2017}, \cite{Vincenzo2021} propose that this prompt contribution comes from the $r$-process production in neutron rich winds from rapidly rotating, highly magnetized winds from proto-neutron stars.  This contribution would be associated with CCSN, but it is not accounted for in the \citetalias{sukhbold2016} calculations used here.  Alternatively, these elements could be produced by the $s$-process in rapidly rotating massive stars \citep{pignatari2008, chiappini2011, frischknecht2012, cescutti2013, Vincenzo2021}, then dispersed in winds or ejected in CCSN explosions.

Appendix~\ref{ap:LC18} repeats our IMF-averaged yield calculations using the CCSN yield set of \citet{lc18}, with Figure~\ref{fig:lc18_fracCCSN} presenting an observational comparison similar to that of Figure~\ref{fig:fracCCSN}. For most elements, we find discrepancies of the same sign but somewhat different magnitudes (sometimes larger and sometimes smaller). Some differences may arise from the much sparser \citet{lc18} mass grid \add{(9 stellar masses vs. 200 in \citetalias{sukhbold2016})}, which does not resolve the complex mass dependence seen in Figure~\ref{fig:mass_yield} and in Figure~\ref{fig:mass_yield_all} below. However, the qualitative similarity of Figures~\ref{fig:fracCCSN} and~\ref{fig:lc18_fracCCSN} suggests that the most significant differences between predicted and empirical yield ratios are robust to the differences in massive star evolution and explosion physics between \citetalias{sukhbold2016} and \citet{lc18}.

\subsection{Interplay with the Stellar IMF}
\label{subsec:disk_imf}

\begin{figure}[!htb]
\begin{centering}
\includegraphics[width=.65\columnwidth]{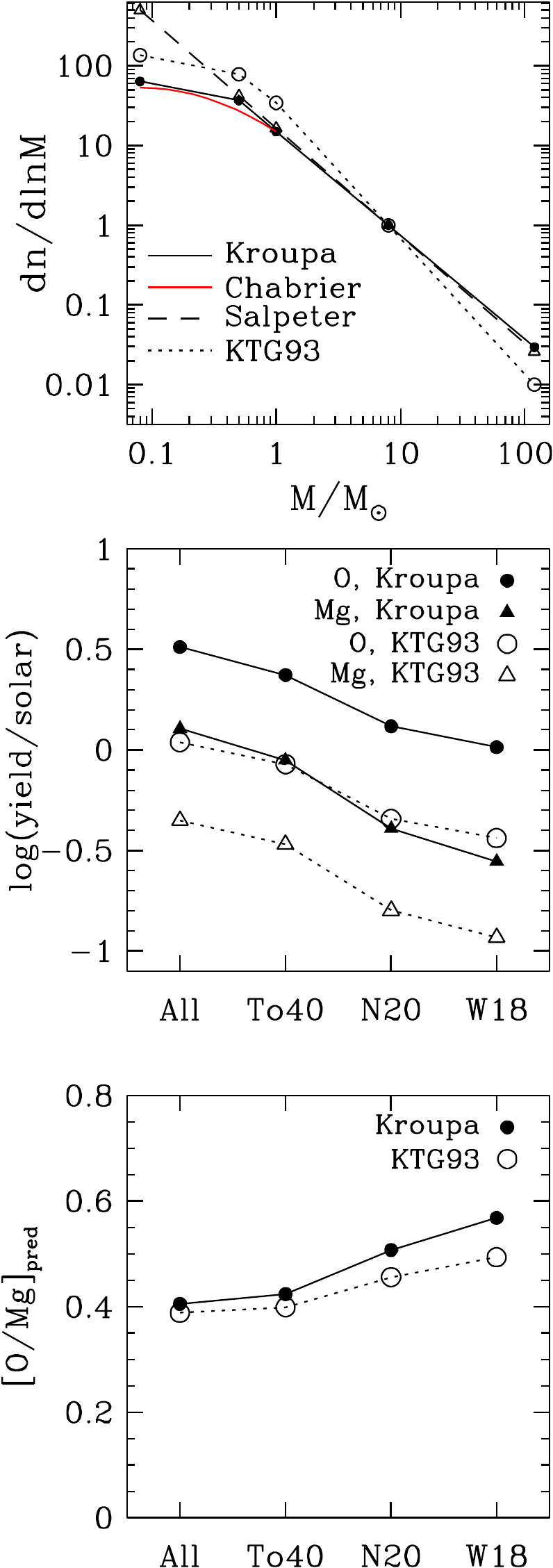}
\caption{({\it Top}) Several forms of the stellar IMF, normalized to equal amplitude at $8\Msun$.  The solid black line shows the \citet{kroupa2001} IMF adopted in our calculations.  The \cite{Chabrier2003} IMF is the same as \cite{kroupa2001} above $1 \Msun$ but follows the log-normal form shown by the red solid curve below $1 \Msun$.  The dashed line shows the \cite{Salpeter1955} IMF, a pure power-law with $dn/d\ln M \propto M^{-1.35}$. The dotted line shows the IMF of \citetalias{Kroupa1993}. Points mark the minimum stellar mass of $0.08 \Msun$ and the locations where either the \citetalias{Kroupa1993} or the \cite{kroupa2001} IMF changes slope.
({\it Middle}) IMF-averaged yields of O (circles) and Mg (triangles) for the Kroupa (filled symbols) and KTG93 (open symbols) IMFs, computed for the All Explode, Explode to 40$\Msun$, Z9.6+N20, and Z9.6+W18 BH formation landscapes.  Yields are normalized to the solar abundances of O and Mg.
\add{({\it Bottom}) [O/Mg] abundances predicted from the IMF-averaged yields in the middle panel.}}
\label{fig:imf}
\end{centering}
\end{figure}

Although it is one of the most basic characteristics of star formation and a crucial input to modeling the chemical evolution and light output of galaxies, the IMF remains uncertain, even in the well explored regime of the Milky Way disk. Figure~\ref{fig:imf} compares the widely used \cite{kroupa2001} IMF, adopted in this study, to several alternatives. We have normalized all of them to the same amplitude at $M=8\Msun$, our assumed minimum mass for a supernova progenitor.  Above $M=0.5\Msun$, the Kroupa IMF has nearly the same slope as the classic \cite{Salpeter1955} IMF, $-1.3$ vs. $-1.35$ in $dn/d\ln M$. However, the Kroupa IMF breaks to a shallower, $-0.3$ power law between $0.08\Msun$ and $0.5 \Msun$, consistent with many indications that the Salpeter extrapolation overpredicts the observed number of low mass disk stars. The \cite{Chabrier2003} IMF, also widely used, has the same power-law slope as \cite{kroupa2001} above $1 \Msun$, but below $1 \Msun$ it follows a log-normal rather than a power-law form.  The Kroupa and Chabrier IMFs are quite similar, despite the different functional forms. The dotted line in Figure~\ref{fig:imf} shows the IMF of \citet*[][KTG93]{Kroupa1993}, which has a power-law slope of $-1.7$ above $1 \Msun$ instead of $-1.3$. Many chemical evolution models (e.g., \citealp{matteucci1989, romano2005, romano2010}; and recently, \citealp{palla2020, spitoni2021}) have adopted either the KTG93 IMF or a similar IMF based on \cite{Scalo1986}.

As shown by Equation~\ref{eq:vice_yield}, the IMF-averaged yield is the ratio of a ``yield integral'' (numerator) to a ``mass integral''  (denominator), with the IMF normalization cancelling out of the ratio. For the Salpeter and Kroupa IMFs, the yield integrals are nearly identical, while the Salpeter mass integral is higher by a factor of 1.55 if the two IMFs are matched at $8 \Msun$.  With {\tt VICE} integrations we find that the IMF-averaged yields of all elements are reduced by a nearly identical factor of 1.6 when changing from a Kroupa IMF to a Salpeter IMF, slightly larger than 1.55 because of the steeper fall-off at $M > 8\Msun$. For the Chabrier IMF matched to Kroupa at $8\Msun$, the yield integrals are identical and the mass integral is nearly identical, so yield predictions are very similar.

For the KTG93 IMF, again matched to the Kroupa IMF at $8\Msun$, the mass integral increases by a factor of 1.66, and the steeper slope above $8 \Msun$ also changes the yield integral by reducing the number of high mass stars.  Figure~\ref{fig:imf} plots the IMF-averaged O yield and Mg yield for the Kroupa and KTG93 IMFs and the four BH landscape scenarios considered in Figure~\ref{fig:xmg} and~\ref{fig:fracCCSN}, All Explode, Explode to 40$\Msun$, Z9.6+N20, and Z9.6+W18. These yields are normalized to the corresponding solar mass fractions $Z_{\rm O}$ and $Z_{\rm Mg}$.  For the All Explode model, changing from the Kroupa IMF to the KTG93 IMF reduces the O and Mg yields by factors of 2.98 and 2.86, respectively. The impact is larger than the factor of 1.66 from the mass integral because of the substantially reduced number of massive stars in the KTG93 IMF.  The reduction is larger for O than for Mg because a larger fraction of O comes from the upper end of the IMF (Figure~\ref{fig:omg_chunk}).  In both scenarios, the yields decline as we go to landscapes with more BH formation. The decline is shallower for the KTG93 IMF, because the stars massive enough to collapse to BHs are more rare, but this is a small effect. For the Z9.6+W18 model, the yield ratios are still 2.83 (O) and 2.38 (Mg) between the two IMFs.

The O/Mg problem discussed in Section~\ref{subsec:OMg} is manifested in Figure~\ref{fig:imf} by the gap between the O and Mg points for a specified choice of IMF and explosion landscape. As discussed below, galactic winds can reduce the abundances of the ISM and newly forming stars below the IMF-averaged yield, but they cannot change the ratio of O to Mg unless they differentially affect these two elements.  It is therefore unlikely that chemical evolution effects can remove this discrepancy in the yield ratios. As previously noted, the O/Mg gap becomes larger for a less explosive supernova landscape. About half of the O overproduction can be attributed to winds, as they continue to contribute O from high mass stars regardless of BH formation. The steeper KTG93 IMF slightly reduces the O/Mg gap, but only slightly.  It is evident from Figure~\ref{fig:imf} that the resolution of this problem does not lie in changing the IMF.

We have investigated the full set of abundance ratios plotted in Figure~\ref{fig:fracCCSN} for the KTG93 IMF and find little qualitative difference, though when models are normalized to reproduce solar Mg, the overproduction of C and O are modestly reduced for the steeper IMF. With the KTG93 IMF, the $\Delta$[O/Mg]$_\text{cc}$ discrepancy falls by $\sim 0.02$ dex for all explosion landscapes. The change in IMF affects C the most, as the yield contribution from high mass stars is lessened under the KTG93 IMF. The steeper IMF decreases $\Delta$[C/Mg]$_\text{cc}$ by 0.11 dex (All Explode) to 0.20 dex (Z9.6+W18). Using the yields of \cite{Chieffi2013} and assuming BH formation above $30 \Msun$, \cite{Griffith2021} found that changes of $\pm 0.3$ in the high-mass IMF slope produce changes of $0.01-0.03$ dex in the [X/Mg] yield ratios for most elements, and 0.05 dex for Ni.  These differences are typically smaller than the differences that we find when comparing different explosion landscapes, at least for scenarios as distinct as Explode to 40 and Z9.6+N20. This comparison suggests that IMF uncertainties are not a major limitation in testing landscape scenarios through yield ratios, and conversely that it is difficult to constrain the IMF through yield ratios unless BH formation is well characterized. 

\add{
\subsection{Interplay with Galactic Winds} \label{subsec:galwinds}}

The impacts of the IMF and BH formation on absolute yields have implications for the role of galactic winds in chemical evolution. For a simple one-zone chemical evolution model with inflow, outflow, and a constant star formation rate (SFR), elemental abundances saturate at an equilibrium value
\begin{equation}
Z_{\rm eq} = y/(1+\eta-r)~,
\label{eqn:zeq}
\end{equation}
where $y$ is the IMF-averaged yield, $\eta = \dot{M}_{\rm out}/{\rm SFR}$ is the outflow mass loading factor, and $r$ is the recycling factor \citep{Weinberg2017}. For a slowly declining SFR the equilibrium abundance is slightly higher, an effect we will ignore in this discussion. The recycling factor
\begin{equation}
r = \frac{\int_{m_{\rm TO}}^{m_u} (m-m_{\rm rem}) \frac{dN}{dm} dm }{\int_{m_l}^{m_u} m \frac{dN }{ dm} dm} 
\label{eqn:recycling}
\end{equation}
is 0.44 and 0.31 for a Kroupa and KTG93 IMF, respectively, where we have adopted a turnoff mass $m_{\rm TO}=1 \Msun$ and approximated the remnant mass as $m_{\rm rem} = 0.6 \Msun$ for $1 < m < 8 \Msun$ and $m_{\rm rem} = 1.5 \Msun$ for $m_{\rm rem} \geq 8 \Msun$. For the All Explode scenario and Kroupa IMF, evolving to solar O abundances requires substantial outflows, as long argued in models of the mass-metallicity relation and Milky Way chemical evolution (e.g., \citealt{Finlator2008,Peeples2011,Zahid2012,andrews2017,rybizki2017,Weinberg2017}).  Figure~\ref{fig:imf} implies $1+\eta-r \approx 3.3$ or $\eta \approx 2.7$.  However, because Mg is underproduced relative to O, reaching solar Mg at equilibrium only requires $1+\eta-r \approx 1.3$, or $\eta \approx 0.75$.  With the Z9.6+W18 explosion landscape, on the other hand, the IMF-averaged O yield is nearly equal to the solar O abundance, implying $1+\eta-r \approx 1$ and a small outflow mass loading $\eta \approx 0.45$.  The Mg yield in this scenario is a factor of 3.6 {\it below} the solar abundance, making it nearly impossible to explain the solar Mg level with this explosion landscape and the Mg yields of the S16 models.

Changing to the KTG93 IMF reduces yields by a factor of $2.5-3$ for any given explosion landscape. Even for the All Explode scenario, the IMF-averaged O yield is about equal to the solar abundance, which is why chemical evolution models adopting this IMF (or that of \citet{Scalo1986}) are generally able to reproduce solar neighborhood abundances with minimal or no outflows \citep[e.g.][]{matteucci1989, romano2010}. For the Z9.6+N20 or Z9.6+W18 landscapes, the predicted O yields are less than half the solar abundance, which would make it difficult to construct viable chemical evolution models. With the KTG93 IMF the predicted Mg yields are well below the solar abundance for all explosion landscapes.

\section{Summary}\label{sec:summary}

In this paper, we present the IMF-averaged CCSN yields from the solar metallicity models of \citetalias{sukhbold2016} for a variety of BH landscapes. We construct a fully exploding yield set by artificially exploding stars that collapse under the Z9.6+W18 engine, with explosion energy and mass cut parameters motivated by the engine driven models. With this yield set, we explore the complex behavior of the explosive and wind yields with progenitor mass, classifying elements as N-like, O-like, Si-like, and Fe-like. From the finely sampled plots of $\MZAMS$ vs. yield (Figures~\ref{fig:mass_yield} and~\ref{fig:mass_yield_all}), we better understand the mechanisms producing each element and the mass ranges that contribute most to their production.

With explosive yields for all stellar models, we can calculate IMF averaged yields for any BH landscape with {\tt VICE}. In this paper we present yields of a landscape where all stars explode, landscapes where stars with masses under some limit explode (e.g. Explode to 40$\Msun$), and landscapes produced in \citetalias{sukhbold2016} (Z9.6+W18 and Z9.6+N20) that have interleaved regions of explosion and regions of implosion. To understand the yields produced by a wider range of landscapes than those in \citetalias{sukhbold2016}, we construct a method for predicting stellar explodability based its correlation with progenitor properties $M_4$ and $M_4\mu_4$ from \citet{ertl2016}. With this method, we create continuous BH formation landscapes for all degrees of explodability. We present the IMF-averaged yields of all landscapes in Figure~\ref{fig:absYield} and the [X/Mg] values of four representative landscapes in Figure~\ref{fig:xmg}. Figure~\ref{fig:fracCCSN} compares the predicted [X/Mg] and [X/O] ratios to the values inferred observationally by combining solar abundances \citep{asplund2009} with empirical decompositions that separate the contributions of CCSN from delayed sources \citep{weinberg2019, griffith19}. Our main findings are:
\begin{itemize}
    \item The Z9.6+W18 and Z9.6+N20 landscapes predict similar abundance yields for most elements, with the additional explodability of N20 slightly increasing the yields. 
    \item For $\alpha$-elements, the absolute yields of the Explode to $40\Msun$ landscape are typically $2-3\times$ higher than those of Z9.6+W18, while going to the All Explode scenario increases yields by a further $20-40\%$.
    \item For C and N, the IMF-averaged yields have only weak dependence on the explosion landscape because they are dominated by the winds from $\MZAMS > 40 \Msun$ stars, which are assumed to escape in all scenarios.
    \item The dependence of relative yields on the explosion landscape is generally much weaker than that of absolute yields, with typical variations of $0.05-0.2$ dex in [X/Mg]
    \item For most elements through the Fe-peak, predicted [X/Mg] ratios decrease for more explosive landscapes. Mn and Ni are notable exceptions to this trend. For $s$-process elements the predicted [X/Mg] ratios are usually well below solar, more so as explodability decreases.
    \item All landscapes overproduce the O/Mg ratio, by factors of $2.5-4$. As shown in Section~\ref{subsec:OMg}, some of this discrepancy arises from stars with $\MZAMS > 30 \Msun$, which contribute significantly to the IMF-averaged yield of O but not to Mg (Figure~\ref{fig:omg_chunk}). However, the O/Mg problem is present even in the lower mass stars.
    \item When normalized to solar Mg, the models strongly overproduce C, Na, and Cu, in addition to O, and the less explosive models (e.g., Z9.6+W18) overproduce S, K, Fe, and Co. When normalized to solar O, all models underpredict Ca, Ti, and V in addition to Mg, and the less explosive models overproduce C and underproduce Mn. The All Explode and Explode to 40$\Msun$ scenarios still overproduce Cu, but the C overproduction is greatly reduced when O is the reference element.
    \item \add{C and N yields are dominated by wind production, causing their absolute yields to be nearly independent of the explosion landscape and the [C/Mg] and [N/Mg] ratios to vary by $\sim 0.6$ dex between the W18 and All Explode landscapes. If the CCSN component of observed C and N abundances can be isolated from that of other production sources (such as AGB stars), both elements could be diagnostic of the Milky Way's BH landscape.}
    \item Mn and Ni/Mg are the most promising diagnostic for distinguishing a simple mass threshold for BH formation from a more complex landscape because they are efficiently produced in compact progenitors that explode in one scenario but not the other (Figure~\ref{fig:delta_xmg}, Figure~\ref{fig:mass_cp}).
    \item Changing from the \citet{kroupa2001} IMF to the \citet{Kroupa1993} IMF, which has a much steeper high mass slope, lowers the predicted yields by a factor of $2-3$ for a given landscape but has little impact on the relative yields (Section~\ref{subsec:disk_imf}).
    
\end{itemize}

Because none of our models achieves good across-the-board agreement with observations, we cannot currently say whether observations favor or disfavor a complex landscape of BH formation as predicted by recent theoretical studies \citep{ugliano2012, pejcha2015, sukhbold2016, ertl2016}. We discuss possible resolutions of some of the most significant discrepancies in Section~\ref{subsec:disc_disrepancies}. For the O/Mg problem, resolution might lie in the triple-$\alpha$ and $^{12}$C($\alpha$,$\gamma$) nuclear cross sections, which affect the ratio of C/O produced during He burning and the amount of $^{12}$C fuel available for Mg production. Overproduction of C and N could be mitigated by sharply reducing the wind mass loss from $\MZAMS > 40\Msun$ stars \textit{and} assuming that these more massive stars collapse to BHs rather than release their accumulated C, N, and O when they explode. Underproduction of V and Mn could be a diagnostic of the core collapse and subsequent bounce, which affects the electron fraction and thus the availability of free neutrons for odd-$Z$ element production in explosive nucleosynthesis. Overproduction of Cu is linked to the availability of free neutrons for $s$-process nucleosynthesis in the He shell, and thus to the rate of $^{22}$Ne($\alpha$, n) reaction that produces these neutrons. Underproduction of Sr, Y, and Zr likely reflects the contribution of other processes to these elements, such as $r$-process nucleosynthesis in proto-neutron star winds \citep[e.g.,][]{Vlasov2017, Vincenzo2021} or $s$-process production in rapidly rotating massive stars \citep[e.g.,][]{frischknecht2012}. Binary star evolution can accelerate stripping of stellar envelopes and thereby change the structure of massive star cores, potentially affecting both BH formation and nucleosynthesis \citep{Patton2020, Laplace2021}.

Some of these discrepancies could be affected by systematic uncertainties in correcting solar abundances for non-CCSN contributions, and for some elements in the solar abundances themselves. However, these observational systematics are moderate and will improve with future data sets. Comparisons like those presented here can then serve as powerful diagnostics of massive star evolution, supernova explosion physics, and black hole formation. 

\section*{Acknowledgments} 

We thank Todd Thompson for illuminating discussions on these topics. We thank the reviewer for their time and comments which have improved the text. This work was supported in part by NSF grant AST-1909841. T.S. was supported by NASA through the NASA Hubble Fellowship grant \#60065868 awarded by the Space Telescope Science Institute, which is operated by the Association of Universities for Research in Astronomy, Inc., for NASA, under contract NAS5-26555. D.H.W acknowledges the hospitality of the Institute for Advanced Study and the support of the W.M. Keck Foundation and the Hendrichs Foundation. F.V. acknowledges the support of a Fellowship from the Center for Cosmology and AstroParticle Physics at the Ohio State University.

\software{Matplotlib \citep{hunter2007}, NumPy \citep{harris2020}, {\tt VICE} \citep{johnson2020}}

{\tt VICE} is an open-source python package available for Linux and Mac OS X. Windows users should install and use {\tt VICE} entirely within the Windows Subsystem for Linux. It can be installed in a terminal via {\tt pip install vice}, after which {\tt vice -{}-docs} will launch a web browser to the documentation at https://vice-astro.readthedocs.io. {\tt vice -{}-tutorial} will also launch a web browser, but to a jupyter notebook in the GitHub repository intended to familiarize first-time users with {\tt VICE}'s API. {\tt VICE} requires Python 3.6 or later.

\appendix

\section{Elemental Yields}\label{ap:yields}

\begin{figure*}[!htb]
 \includegraphics[width=.96\textwidth]{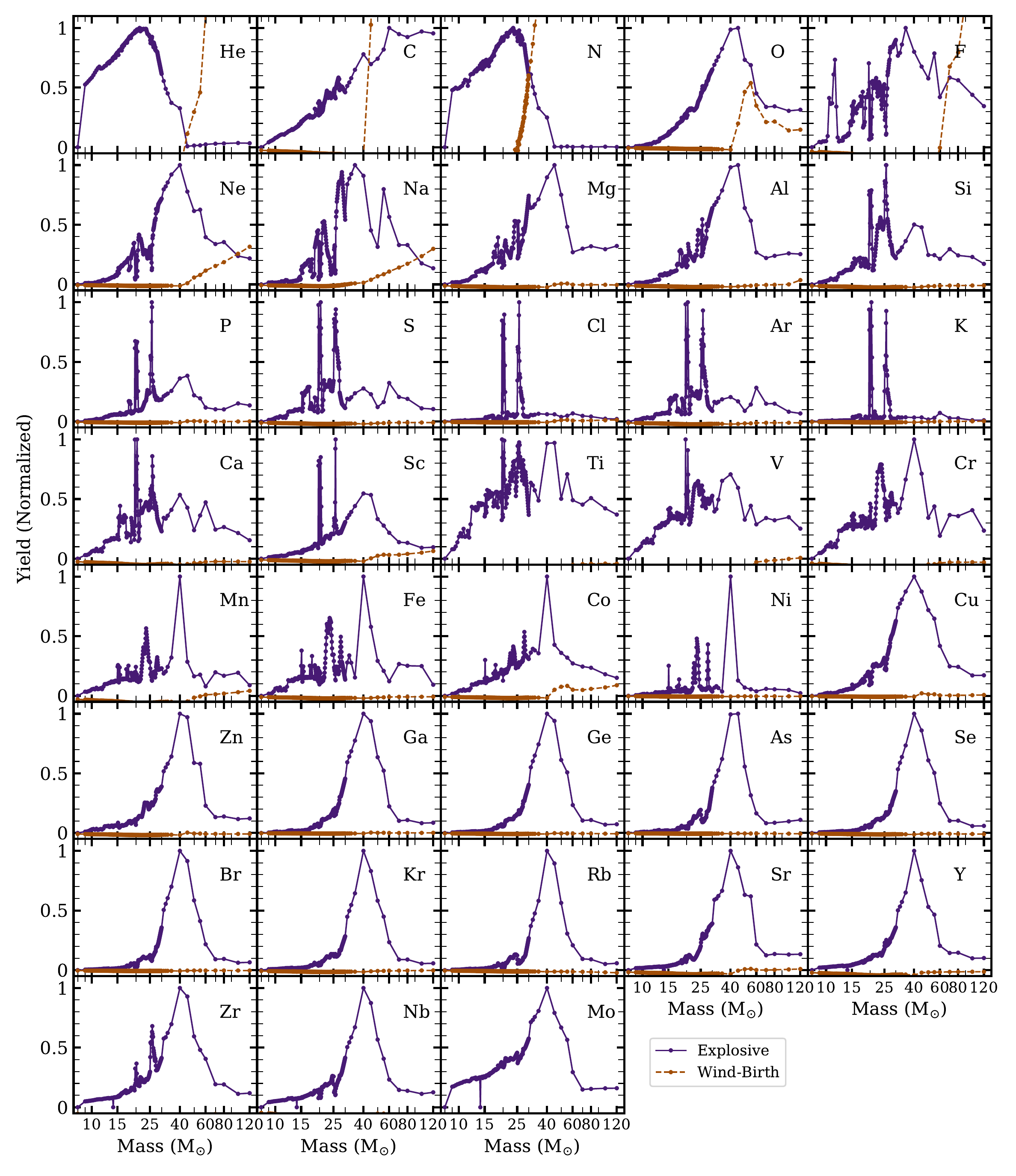}
 \caption{Explosive (dark purple, solid line) and net wind (dark orange, dashed line) yields in $\Msun$ produced per star of a given progenitor mass for all elements in \citetalias{sukhbold2016}. Net wind yields are calculated by subtracting the elemental birth abundances per star from the reported wind contributions. All yields have been normalized such that the maximum explosive yield is 1.}
 \label{fig:mass_yield_all}
\end{figure*}

Figure~\ref{fig:mass_yield_all} shows the explosive and net wind yields for all elements in \citetalias{sukhbold2016}, similar to Figure~\ref{fig:mass_yield}. All yields have been normalized so that the maximum explosive yield is 1. The net wind yields of He, C, N, and F continue on the shown trajectory for the highest mass stars. For elements such as Ti, V, Cr, Zr, Nb, and Mo, the net wind yields are less than zero and are not shown.

As discussed in Section~\ref{subsec:yield_v_mass}, the elemental yield vs. $\MZAMS$ trends can be categorized as N-like (He), O-like (Ne, Mg, Al, Cu, Zn), Si-like (P, S, Cl, Ar, K, Ca, Sc), and Fe-like (V, Cr, Mn, Co, Ni). Neutron capture elements (Ga, Ge, As, Se, Br, Kr, Rb, Sr, Y, Zr, Nb, and Mo) all also resemble O, as the yields climb to a peak at 40$\Msun$, then decrease at higher masses because of their formation in He burning shell.

The IMF-averaged yields for all elements included in Figure~\ref{fig:mass_yield_all} can be found in Table~\ref{tab:yield}. Here we provide the yields for the All Explode, Explode to 40$\Msun$, Explode to 21$\Msun$, Z9.6+N20, and Z9.6+W18 landscapes. Yields for other upper mass or $e_0$ landscapes are available upon request or can be calculated with {\tt VICE}. Abundance ratios, such as those plotted in Figure~\ref{fig:xmg}, can be calculated from the IMF-averaged yields in Table~\ref{tab:yield} and Equation~\ref{eq:xy}.

\begin{table}
\caption{Net yields in $\Msun$ per $\Msun$ formed (including explosive and wind contributions) for all elements included in \citetalias{sukhbold2016}. We report yields for five models, All explode, Explode to 40 $\Msun$, Explode to 21 $\Msun$ (mass cut where the Mg yield equals the W18 Mg yield), N20 landscape, and W18 landscape. \label{tab:yield}}
\centering
\begin{tabular}{lrrrrr}
\toprule
Element &   All Exp &  To 40 $\Msun$&  To 21 $\Msun$ &       N20 &       W18 \\
\hline
     He &  1.14e-02 &   1.11e-02 &   4.18e-03 &  4.15e-03 &  2.33e-03 \\
      C &  7.11e-03 &   6.41e-03 &   5.69e-03 &  6.05e-03 &  5.65e-03 \\
      N &  4.10e-04 &   4.08e-04 &   3.47e-04 &  3.48e-04 &  3.33e-04 \\
      O &  1.81e-02 &   1.31e-02 &   5.29e-03 &  7.29e-03 &  5.75e-03 \\
      F &  3.18e-07 &   2.62e-07 &   1.34e-07 &  1.51e-07 &  1.27e-07 \\
     Ne &  3.34e-03 &   2.43e-03 &   7.75e-04 &  1.13e-03 &  8.18e-04 \\
     Na &  8.18e-05 &   6.27e-05 &   1.86e-05 &  2.92e-05 &  2.12e-05 \\
     Mg &  8.78e-04 &   6.11e-04 &   1.90e-04 &  2.80e-04 &  1.91e-04 \\
     Al &  8.61e-05 &   5.92e-05 &   1.42e-05 &  2.31e-05 &  1.54e-05 \\
     Si &  7.83e-04 &   6.23e-04 &   2.50e-04 &  3.54e-04 &  2.72e-04 \\
      P &  1.13e-05 &   8.67e-06 &   3.38e-06 &  5.08e-06 &  4.03e-06 \\
      S &  4.12e-04 &   3.44e-04 &   1.65e-04 &  2.42e-04 &  1.93e-04 \\
     Cl &  4.91e-06 &   4.30e-06 &   1.95e-06 &  3.12e-06 &  2.75e-06 \\
     Ar &  7.19e-05 &   5.94e-05 &   2.70e-05 &  4.00e-05 &  3.18e-05 \\
      K &  3.39e-06 &   3.05e-06 &   1.53e-06 &  2.37e-06 &  2.13e-06 \\
     Ca &  3.96e-05 &   3.14e-05 &   1.39e-05 &  1.91e-05 &  1.43e-05 \\
     Sc &  3.54e-08 &   2.56e-08 &   8.11e-09 &  1.23e-08 &  9.48e-09 \\
     Ti &  1.45e-06 &   1.15e-06 &   5.87e-07 &  7.92e-07 &  5.73e-07 \\
      V &  1.20e-07 &   9.49e-08 &   3.77e-08 &  4.98e-08 &  3.28e-08 \\
     Cr &  8.75e-06 &   6.80e-06 &   2.65e-06 &  3.45e-06 &  2.32e-06 \\
     Mn &  4.19e-06 &   3.12e-06 &   8.23e-07 &  9.67e-07 &  5.35e-07 \\
     Fe &  1.58e-03 &   1.20e-03 &   5.29e-04 &  5.99e-04 &  4.50e-04 \\
     Co &  5.52e-06 &   4.15e-06 &   1.80e-06 &  2.28e-06 &  1.76e-06 \\
     Ni &  1.19e-04 &   8.68e-05 &   2.40e-05 &  2.55e-05 &  1.86e-05 \\
     Cu &  3.59e-06 &   2.45e-06 &   5.79e-07 &  9.98e-07 &  7.04e-07 \\
     Zn &  2.97e-06 &   1.91e-06 &   3.97e-07 &  7.11e-07 &  4.86e-07 \\
     Ga &  3.05e-07 &   1.91e-07 &   2.27e-08 &  5.24e-08 &  3.46e-08 \\
     Ge &  6.19e-07 &   3.73e-07 &   2.89e-08 &  8.81e-08 &  5.38e-08 \\
     As &  3.73e-08 &   2.20e-08 &   2.35e-09 &  5.58e-09 &  3.38e-09 \\
     Se &  2.44e-07 &   1.44e-07 &   6.05e-09 &  2.89e-08 &  1.62e-08 \\
     Br &  5.91e-08 &   3.45e-08 &   2.05e-09 &  7.06e-09 &  4.07e-09 \\
     Kr &  1.42e-07 &   8.05e-08 &   1.62e-09 &  1.39e-08 &  6.91e-09 \\
     Rb &  2.60e-08 &   1.39e-08 &  -6.93e-11 &  1.91e-09 &  7.05e-10 \\
     Sr &  3.30e-08 &   1.91e-08 &  -4.72e-10 &  3.55e-09 &  1.15e-09 \\
      Y &  4.55e-09 &   2.57e-09 &  -2.60e-10 &  2.55e-10 & -1.01e-10 \\
     Zr &  4.72e-09 &   2.41e-09 &  -1.24e-09 & -3.37e-10 & -8.54e-10 \\
     Nb &  3.12e-10 &   1.51e-10 &  -8.84e-11 & -5.30e-11 & -8.65e-11 \\
     Mo & -6.20e-11 &  -2.02e-10 &  -4.95e-10 & -4.59e-10 & -5.12e-10 \\
\hline
\end{tabular}
\end{table}

\add{
\section{The Two-Process Model}\label{ap:2proc}}

\add{This is a new appendix. We have moved the black text from Section 5 to this appendix and have added the green text. -EJG}

\add{The two-process model, developed by \citet{weinberg2019}, quantifies the relative CCSN and SNIa contribution to elements and describes each element's CCSN and SNIa metallicity dependence. The model describes a star's abundances as the sum of a CCSN and SNIa process ($\pcc$ and $\pIa$), with amplitudes $\Acc$ and $\AIa$. While the processes are fixed for each element, $\Acc$ and $\AIa$ vary between stars. \citet{weinberg2019} express the CCSN and SNIa processes as a power law with slopes $\acc$ and $\aIa$ such that
\begin{equation}\label{eq:pcc}
    \pcc(Z) = \pccsun \cdot 10^{\acc\mgh}
\end{equation}
and 
\begin{equation}\label{eq:pIa}
    \pIa(Z) = \pIasun\cdot 10^{\aIa\mgh}
\end{equation}
where $\pccsun$ and $\pIasun$ represent the contribution of each process at $\mgh=0$. The process contributions at solar $\mgh$ define the ratio of the processes for an element X:
\begin{equation}\label{eq:RIa}
    \RIa = \frac{\pIasun}{\pccsun}.
\end{equation}
This can also be expressed as the fractional CCSN contribution to element X at $\mgh=\feh=0$, where
\begin{equation}\label{eq:fcc}
    \fcc = \frac{1}{1+\RIa}.
\end{equation}}

\add{In the main text, we use results from the two-process model fit to GALAH \citep{griffith19} and APOGEE \citep{weinberg2019} elements to contrast theoretical CCSN yields with empirical CCSN yields.} Figure~\ref{fig:fracCCSN} shows the ratio of the BH landscapes' CCSN yields of element X to that of Mg (top) and O (bottom), divided by the ratio of estimated CCSN yield to that of Mg from \citet{griffith19}. We plot the $\log_{10}$ of this ratio to show the difference between the theoretical and empirical CCSN [X/Mg] or [X/O]. \add{The values $\Delta [\text{X}/\text{Mg}]_{\text{cc}}$ and $\Delta [\text{X}/\text{Mg}]_{\text{cc}}$ can be expressed in terms of the theoretical yield ($\text{Y}_{\text{cc}}^{\text{X}}$) and the CCSN process amplitude (Equation~\ref{eq:pcc}) such that}
\begin{equation} \label{eq:delta_mg}
    \Delta [\text{X}/\text{Mg}]_{\text{cc}} = \log_{10} \bigg( \frac{\text{Y}^{\text{X}}_{\text{cc}}}{\text{Y}^{\text{Mg}}_{\text{cc}}} \div \frac{p^{\text{X}}_{\text{cc}}}{p^{\text{Mg}}_{\text{cc}}} \bigg)
\end{equation}
and 
\begin{equation}\label{eq:delta_o}
    \Delta [\text{X}/\text{O}]_{\text{cc}} = \log_{10} \bigg( \frac{\text{Y}^{\text{X}}_{\text{cc}}}{\text{Y}^{\text{O}}_{\text{cc}}} \div \frac{p^{\text{X}}_{\text{cc}}}{p^{\text{Mg}}_{\text{cc}}} \bigg),
\end{equation}
Both $\Delta [\text{X}/\text{Mg}]_{\text{cc}}$ and $\Delta [\text{X}/\text{O}]_{\text{cc}}$ contain the term $p_{\text{cc}}^{\text{Mg}}$ as both are based on the two-process decomposition that assumes Mg is purely produced in CCSN.
An equivalent and perhaps more intuitive form of these equations is
\begin{equation}\label{eq:delta_mg_b}
    \Delta [\text{X}/\text{Mg}]_{\text{cc}} = \log_{10} \frac{\fcc^{\text{X}} \text{Y}_{\text{cc}}^{\text{X}}/(\text{X}/\text{H})_{\odot}}{\fcc^{\text{Mg}} \text{Y}_{\text{cc}}^{\text{Mg}}/(\text{Mg}/\text{H})_{\odot}}
\end{equation}
and
\begin{equation}\label{eq:delta_o_b}
    \Delta [\text{X}/\text{O}]_{\text{cc}} = \log_{10} \frac{\fcc^{\text{X}} \text{Y}_{\text{cc}}^{\text{X}}/(\text{X}/\text{H})_{\odot}}{\fcc^{\text{O}} \text{Y}_{\text{cc}}^{\text{O}}/(\text{O}/\text{H})_{\odot}}.
\end{equation}
In \citet{griffith19} we set $\fcc^{\text{Mg}} = 1.0$ by assumption, and we infer $\fcc^{\text{O}} = 1.0$ empirically.

\section{A Comparison to Yields from LC18}\label{ap:LC18}

The main body of this paper has focused on the yields from the neutrino driven explosion models of \citetalias{sukhbold2016}. In this appendix, we apply our analysis to the solar metallicity yields of \citet[][hereafter LC18]{lc18}. Unlike \citetalias{sukhbold2016}, the CCSN explosions of \citetalias{lc18} are driven by a piston of fixed energy. Under this model all stars explode if enough energy is imparted, and a BH landscape is not produced. \citetalias{lc18} present multiple sets of stellar yields, with their recommended set (Set R) including explosive yields of stars with $\MZAMS<25\Msun$ and wind yields of all stars. In this section, we adopt their Set M yields, which include explosive and wind yields for stars of all $\MZAMS$. \citetalias{lc18} report yields for species up to $^{209}$Bi. In this analysis, we use the fully decayed yields. \citetalias{lc18} report yields for 9 stellar masses between 13 and 120$\Msun$ -- a factor of $\sim20$ fewer stellar models that \citetalias{sukhbold2016}. The sparser grid of progenitor masses suggests that the yields of \citetalias{lc18} will be less sensitive to the complex dependence of yield on $\MZAMS$ (as explained in Section~\ref{subsec:yield_v_mass}).

\begin{figure*}[!htb]
 \includegraphics[width=.96\textwidth]{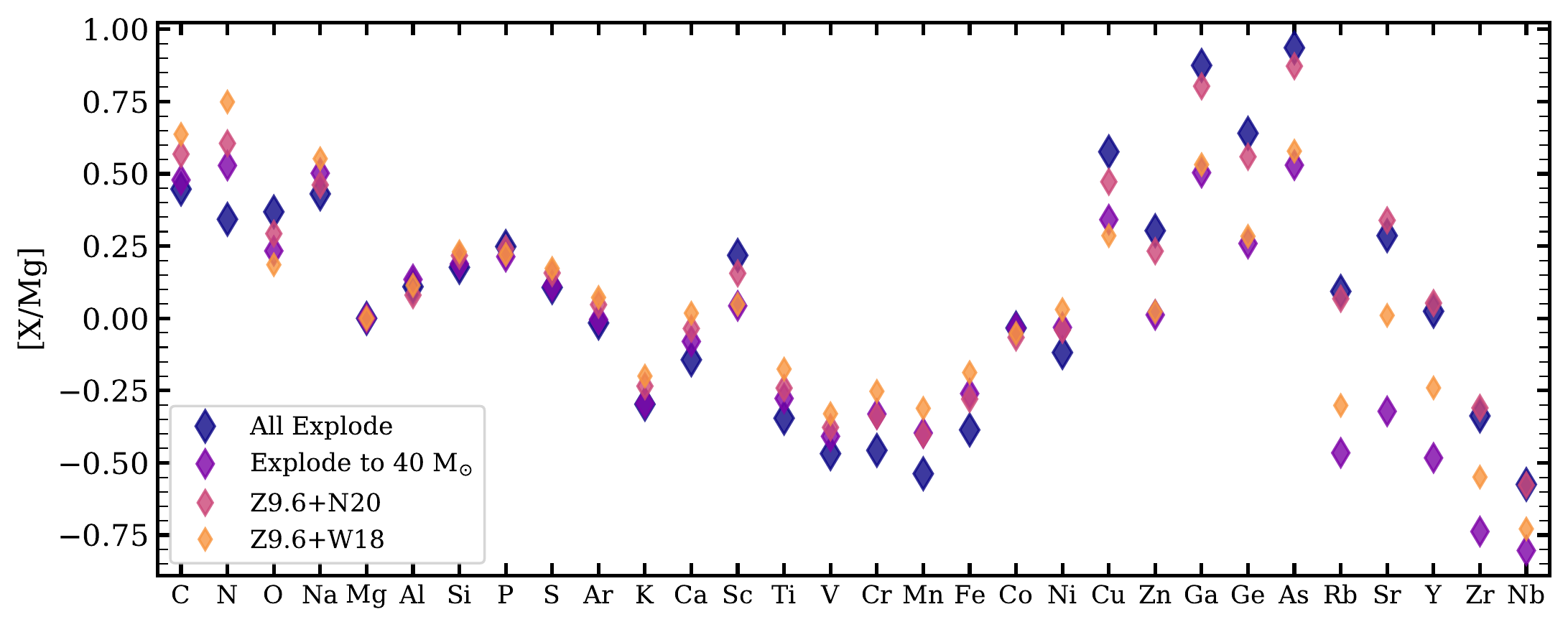}
 \caption{Same as Figure~\ref{fig:xmg}, but for the net, fully decayed Set M yields from \citet{lc18} \add{that provide explosive yields for stars of all masses.}}
 \label{fig:lc18_xmg}
\end{figure*}

With the Set M yields of \citetalias{lc18} we are able to reconstruct the All Explode, Explode to 40$\Msun$, Z9.6+N20, and Z9.6+W18 landscapes by excluding the explosive yields of stars that collapse to BHs. We calculate net IMF-averaged yields and [X/Mg] abundances as described in Sections~\ref{sec:IMF_int} and~\ref{sec:ab_ratios} and plot the resulting [X/Mg] values for the four landscapes in Figure~\ref{fig:lc18_xmg}. Comparing to Figure~\ref{fig:xmg}, we find systematically lower [O/Mg] abundance ratios in \citetalias{lc18} with no clear abundance dependence on BH landscape. While most light-$Z$ and Fe-peak elements show similar behavior between the two yield sets, we find [K/Mg] abundances that are $0.25$ to $0.5$ dex below the \citetalias{sukhbold2016} values for all explosion landscapes, and [Sc/Mg] abundances that are $\sim0.25$ dex above the \citetalias{sukhbold2016} values. Mn follows a similar trend to the other Fe-peak elements in \citetalias{lc18}, in contrast with the reversed dependence on the explodability landscape seen in Figure~\ref{fig:xmg}.

\begin{figure*}[!htb]
 \includegraphics[width=.96\textwidth]{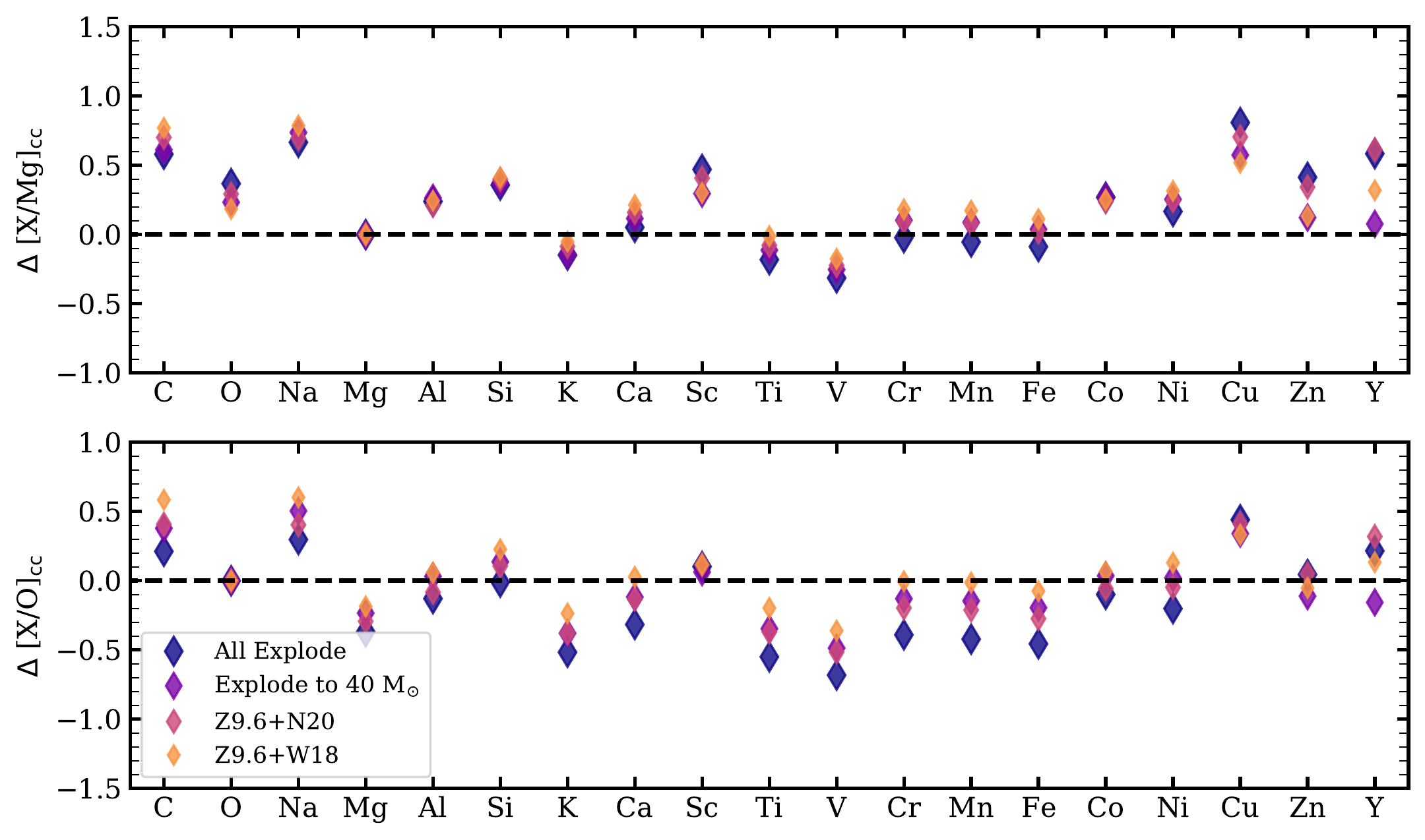}
 \caption{Same as Figure~\ref{fig:fracCCSN} but for the net, fully decayed Set M yields from \citet{lc18} \add{that provide explosive yields for stars of all masses.}}
 \label{fig:lc18_fracCCSN}
\end{figure*}

As in Section~\ref{sec:ab_ratios}, we plot $\Delta$[X/Mg]$_{\rm cc}$ and $\Delta$[X/O]$_{\rm cc}$ (Equations~\ref{eq:delta_mg} and ~\ref{eq:delta_o}) for the \citetalias{lc18} yields in Figure~\ref{fig:lc18_fracCCSN} as a comparison of theoretical and observational CCSN yields. Empirical yield results are again taken from GALAH \citep[][all elements but S]{griffith19} and APOGEE \citep[][S]{weinberg2019}. Many of the differences noted above propagate forward. We see overproduction of C, O, Na, and Cu with respect to Mg as in Figure~\ref{fig:fracCCSN}, though the C overproduction drops by up to $\sim0.5$ dex for the less explosive landscapes. The \citetalias{lc18} yields produce $\sim0.2$ dex less O overproduction than \citetalias{sukhbold2016} and show less dependence on BH landscape. This may be due to the different mass loss/wind treatment in \citet{lc18}, as they attribute far less O to the wind component. \citetalias{lc18} also show increased overproduction of Sc relative to \citetalias{sukhbold2016}.  We see reasonable agreement with the observational yield constraints for Ca, Ti, and Fe-peak elements. While more explosive landscapes are in better agreement with observational constraints for the Fe-peak, this is not seen for all elements. 

Changing our reference from O to Mg decreases the C, Na, and Cu overproduction, but implies the underproduction of Mg, K, Ti, and V. In the $\Delta$[X/O]$_{\rm cc}$ panel of Figure~\ref{fig:lc18_fracCCSN}, we see that the less explosive landscapes tend to be in better agreement with empirical results, in contrast to the top panel. While much of the discussion from Section~\ref{subsec:disc_disrepancies} can inform the features seen in Figure~\ref{fig:lc18_fracCCSN}, a deeper understanding of the discrepancies between the \citetalias{lc18} yields and observational results would require a detailed analysis of the theoretical yields that is beyond the scope of this work.

\bibliography{bibliography} 

\begin{thebibliography}{}
\expandafter\ifx\csname natexlab\endcsname\relax\def\natexlab#1{#1}\fi
\providecommand{\url}[1]{\href{#1}{#1}}
\providecommand{\dodoi}[1]{doi:~\href{http://doi.org/#1}{\nolinkurl{#1}}}
\providecommand{\doeprint}[1]{\href{http://ascl.net/#1}{\nolinkurl{http://ascl.net/#1}}}
\providecommand{\doarXiv}[1]{\href{https://arxiv.org/abs/#1}{\nolinkurl{https://arxiv.org/abs/#1}}}

\bibitem[{{Adams} {et~al.}(2017){Adams}, {Kochanek}, {Gerke}, {Stanek}, \&
  {Dai}}]{adams2017}
{Adams}, S.~M., {Kochanek}, C.~S., {Gerke}, J.~R., {Stanek}, K.~Z., \& {Dai},
  X. 2017, \mnras, 468, 4968, \dodoi{10.1093/mnras/stx816}

\bibitem[{{Amarsi} {et~al.}(2015){Amarsi}, {Asplund}, {Collet}, \&
  {Leenaarts}}]{amarsi2015}
{Amarsi}, A.~M., {Asplund}, M., {Collet}, R., \& {Leenaarts}, J. 2015, \mnras,
  454, L11, \dodoi{10.1093/mnrasl/slv122}

\bibitem[{{Andrews} {et~al.}(2017){Andrews}, {Weinberg}, {Sch{\"o}nrich}, \&
  {Johnson}}]{andrews2017}
{Andrews}, B.~H., {Weinberg}, D.~H., {Sch{\"o}nrich}, R., \& {Johnson}, J.~A.
  2017, \apj, 835, 224, \dodoi{10.3847/1538-4357/835/2/224}

\bibitem[{{Asplund} {et~al.}(2009){Asplund}, {Grevesse}, {Sauval}, \&
  {Scott}}]{asplund2009}
{Asplund}, M., {Grevesse}, N., {Sauval}, A.~J., \& {Scott}, P. 2009, \araa, 47,
  481, \dodoi{10.1146/annurev.astro.46.060407.145222}

\bibitem[{{Beasor} \& {Davies}(2018)}]{beasor2018}
{Beasor}, E.~R., \& {Davies}, B. 2018, \mnras, 475, 55,
  \dodoi{10.1093/mnras/stx3174}

\bibitem[{{Bensby} {et~al.}(2014){Bensby}, {Feltzing}, \& {Oey}}]{bensby2014}
{Bensby}, T., {Feltzing}, S., \& {Oey}, M.~S. 2014, \aap, 562, A71,
  \dodoi{10.1051/0004-6361/201322631}

\bibitem[{{Blanton} {et~al.}(2017){Blanton}, {Bershady}, {Abolfathi},
  {Albareti}, {Allende Prieto}, {Almeida}, {Alonso-Garc{\'\i}a}, {Anders},
  {Anderson}, {Andrews}, {Aquino-Ort{\'\i}z}, {Arag{\'o}n-Salamanca},
  {Argudo-Fern{\'a}ndez}, {Armengaud}, {Aubourg}, {Avila-Reese}, {Badenes},
  {Bailey}, {Barger}, {Barrera-Ballesteros}, {Bartosz}, {Bates}, {Baumgarten},
  {Bautista}, {Beaton}, {Beers}, {Belfiore}, {Bender}, {Berlind}, {Bernardi},
  {Beutler}, {Bird}, {Bizyaev}, {Blanc}, {Blomqvist}, {Bolton}, {Boquien},
  {Borissova}, {van den Bosch}, {Bovy}, {Brandt}, {Brinkmann}, {Brownstein},
  {Bundy}, {Burgasser}, {Burtin}, {Busca}, {Cappellari}, {Delgado Carigi},
  {Carlberg}, {Carnero Rosell}, {Carrera}, {Chanover}, {Cherinka}, {Cheung},
  {G{\'o}mez Maqueo Chew}, {Chiappini}, {Choi}, {Chojnowski}, {Chuang},
  {Chung}, {Cirolini}, {Clerc}, {Cohen}, {Comparat}, {da Costa}, {Cousinou},
  {Covey}, {Crane}, {Croft}, {Cruz-Gonzalez}, {Garrido Cuadra}, {Cunha},
  {Damke}, {Darling}, {Davies}, {Dawson}, {de la Macorra}, {Dell'Agli}, {De
  Lee}, {Delubac}, {Di Mille}, {Diamond-Stanic}, {Cano-D{\'\i}az}, {Donor},
  {Downes}, {Drory}, {du Mas des Bourboux}, {Duckworth}, {Dwelly}, {Dyer},
  {Ebelke}, {Eigenbrot}, {Eisenstein}, {Emsellem}, {Eracleous}, {Escoffier},
  {Evans}, {Fan}, {Fern{\'a}ndez-Alvar}, {Fernandez-Trincado}, {Feuillet},
  {Finoguenov}, {Fleming}, {Font-Ribera}, {Fredrickson}, {Freischlad},
  {Frinchaboy}, {Fuentes}, {Galbany}, {Garcia-Dias},
  {Garc{\'\i}a-Hern{\'a}ndez}, {Gaulme}, {Geisler}, {Gelfand},
  {Gil-Mar{\'\i}n}, {Gillespie}, {Goddard}, {Gonzalez-Perez}, {Grabowski},
  {Green}, {Grier}, {Gunn}, {Guo}, {Guy}, {Hagen}, {Hahn}, {Hall}, {Harding},
  {Hasselquist}, {Hawley}, {Hearty}, {Gonzalez Hern{\'a}ndez}, {Ho}, {Hogg},
  {Holley-Bockelmann}, {Holtzman}, {Holzer}, {Huehnerhoff}, {Hutchinson},
  {Hwang}, {Ibarra-Medel}, {da Silva Ilha}, {Ivans}, {Ivory}, {Jackson},
  {Jensen}, {Johnson}, {Jones}, {J{\"o}nsson}, {Jullo}, {Kamble}, {Kinemuchi},
  {Kirkby}, {Kitaura}, {Klaene}, {Knapp}, {Kneib}, {Kollmeier}, {Lacerna},
  {Lane}, {Lang}, {Law}, {Lazarz}, {Lee}, {Le Goff}, {Liang}, {Li}, {Li},
  {Lian}, {Lima}, {Lin}, {Lin}, {Bertran de Lis}, {Liu}, {de Icaza Lizaola},
  {Long}, {Lucatello}, {Lundgren}, {MacDonald}, {Deconto Machado}, {MacLeod},
  {Mahadevan}, {Geimba Maia}, {Maiolino}, {Majewski}, {Malanushenko},
  {Malanushenko}, {Manchado}, {Mao}, {Maraston}, {Marques-Chaves}, {Masseron},
  {Masters}, {McBride}, {McDermid}, {McGrath}, {McGreer}, {Medina Pe{\~n}a},
  {Melendez}, {Merloni}, {Merrifield}, {Meszaros}, {Meza}, {Minchev},
  {Minniti}, {Miyaji}, {More}, {Mulchaey}, {M{\"u}ller-S{\'a}nchez}, {Muna},
  {Munoz}, {Myers}, {Nair}, {Nandra}, {Correa do Nascimento}, {Negrete},
  {Ness}, {Newman}, {Nichol}, {Nidever}, {Nitschelm}, {Ntelis}, {O'Connell},
  {Oelkers}, {Oravetz}, {Oravetz}, {Pace}, {Padilla}, {Palanque-Delabrouille},
  {Alonso Palicio}, {Pan}, {Parejko}, {Parikh}, {P{\^a}ris}, {Park}, {Patten},
  {Peirani}, {Pellejero-Ibanez}, {Penny}, {Percival}, {Perez-Fournon},
  {Petitjean}, {Pieri}, {Pinsonneault}, {Pisani}, {Poleski}, {Prada},
  {Prakash}, {Queiroz}, {Raddick}, {Raichoor}, {Barboza Rembold}, {Richstein},
  {Riffel}, {Riffel}, {Rix}, {Robin}, {Rockosi}, {Rodr{\'\i}guez-Torres},
  {Roman-Lopes}, {Rom{\'a}n-Z{\'u}{\~n}iga}, {Rosado}, {Ross}, {Rossi}, {Ruan},
  {Ruggeri}, {Rykoff}, {Salazar-Albornoz}, {Salvato}, {S{\'a}nchez}, {Aguado},
  {S{\'a}nchez-Gallego}, {Santana}, {Santiago}, {Sayres}, {Schiavon}, {da Silva
  Schimoia}, {Schlafly}, {Schlegel}, {Schneider}, {Schultheis}, {Schuster},
  {Schwope}, {Seo}, {Shao}, {Shen}, {Shetrone}, {Shull}, {Simon}, {Skinner},
  {Skrutskie}, {Slosar}, {Smith}, {Sobeck}, {Sobreira}, {Somers}, {Souto},
  {Stark}, {Stassun}, {Stauffer}, {Steinmetz}, {Storchi-Bergmann},
  {Streblyanska}, {Stringfellow}, {Su{\'a}rez}, {Sun}, {Suzuki}, {Szigeti},
  {Taghizadeh-Popp}, {Tang}, {Tao}, {Tayar}, {Tembe}, {Teske}, {Thakar},
  {Thomas}, {Thompson}, {Tinker}, {Tissera}, {Tojeiro}, {Hernandez Toledo}, {de
  la Torre}, {Tremonti}, {Troup}, {Valenzuela}, {Martinez Valpuesta},
  {Vargas-Gonz{\'a}lez}, {Vargas-Maga{\~n}a}, {Vazquez}, {Villanova}, {Vivek},
  {Vogt}, {Wake}, {Walterbos}, {Wang}, {Weaver}, {Weijmans}, {Weinberg},
  {Westfall}, {Whelan}, {Wild}, {Wilson}, {Wood-Vasey}, {Wylezalek}, {Xiao},
  {Yan}, {Yang}, {Ybarra}, {Y{\`e}che}, {Zakamska}, {Zamora}, {Zarrouk},
  {Zasowski}, {Zhang}, {Zhao}, {Zheng}, {Zheng}, {Zhou}, {Zhou}, {Zhu},
  {Zoccali}, \& {Zou}}]{blanton2017}
{Blanton}, M.~R., {Bershady}, M.~A., {Abolfathi}, B., {et~al.} 2017, \aj, 154,
  28, \dodoi{10.3847/1538-3881/aa7567}

\bibitem[{{Brown} \& {Woosley}(2013)}]{brown2013}
{Brown}, J.~M., \& {Woosley}, S.~E. 2013, \apj, 769, 99,
  \dodoi{10.1088/0004-637X/769/2/99}

\bibitem[{{Buchmann}(1996)}]{Buc96}
{Buchmann}, L. 1996, \apjl, 468, L127, \dodoi{10.1086/310240}

\bibitem[{{Caffau} {et~al.}(2008){Caffau}, {Ludwig}, {Steffen}, {Ayres},
  {Bonifacio}, {Cayrel}, {Freytag}, \& {Plez}}]{caffau2008}
{Caffau}, E., {Ludwig}, H.~G., {Steffen}, M., {et~al.} 2008, \aap, 488, 1031,
  \dodoi{10.1051/0004-6361:200809885}

\bibitem[{{Caughlan} \& {Fowler}(1988)}]{CF88}
{Caughlan}, G.~R., \& {Fowler}, W.~A. 1988, Atomic Data and Nuclear Data
  Tables, 40, 283, \dodoi{10.1016/0092-640X(88)90009-5}

\bibitem[{{Cescutti} {et~al.}(2013){Cescutti}, {Chiappini}, {Hirschi},
  {Meynet}, \& {Frischknecht}}]{cescutti2013}
{Cescutti}, G., {Chiappini}, C., {Hirschi}, R., {Meynet}, G., \&
  {Frischknecht}, U. 2013, \aap, 553, A51, \dodoi{10.1051/0004-6361/201220809}

\bibitem[{{Chabrier}(2003)}]{Chabrier2003}
{Chabrier}, G. 2003, \pasp, 115, 763, \dodoi{10.1086/376392}

\bibitem[{{Chaplin} {et~al.}(2020){Chaplin}, {Serenelli}, {Miglio}, {Morel},
  {Mackereth}, {Vincenzo}, {Kjeldsen}, {Basu}, {Ball}, {Stokholm}, {Verma},
  {Mosumgaard}, {Silva Aguirre}, {Mazumdar}, {Ranadive}, {Antia}, {Lebreton},
  {Ong}, {Appourchaux}, {Bedding}, {Christensen-Dalsgaard}, {Creevey},
  {Garc{\'\i}a}, {Handberg}, {Huber}, {Kawaler}, {Lund}, {Metcalfe}, {Stassun},
  {Bazot}, {Beck}, {Bell}, {Bergemann}, {Buzasi}, {Benomar}, {Bossini},
  {Bugnet}, {Campante}, {Orhan}, {Corsaro}, {Gonz{\'a}lez-Cuesta}, {Davies},
  {Di Mauro}, {Egeland}, {Elsworth}, {Gaulme}, {Ghasemi}, {Guo}, {Hall},
  {Hasanzadeh}, {Hekker}, {Howe}, {Jenkins}, {Jim{\'e}nez}, {Kiefer},
  {Kuszlewicz}, {Kallinger}, {Latham}, {Lundkvist}, {Mathur}, {Montalb{\'a}n},
  {Mosser}, {Bed{\'o}n}, {Nielsen}, {{\"O}rtel}, {Rendle}, {Ricker},
  {Rodrigues}, {Roxburgh}, {Safari}, {Schofield}, {Seager}, {Smalley},
  {Stello}, {Szab{\'o}}, {Tayar}, {Theme{\ss}l}, {Thomas}, {Vanderspek}, {van
  Rossem}, {Vrard}, {Weiss}, {White}, {Winn}, \& {Y{\i}ld{\i}z}}]{chaplin2020}
{Chaplin}, W.~J., {Serenelli}, A.~M., {Miglio}, A., {et~al.} 2020, Nature
  Astronomy, 4, 382, \dodoi{10.1038/s41550-019-0975-9}

\bibitem[{{Chiappini} {et~al.}(2011){Chiappini}, {Frischknecht}, {Meynet},
  {Hirschi}, {Barbuy}, {Pignatari}, {Decressin}, \& {Maeder}}]{chiappini2011}
{Chiappini}, C., {Frischknecht}, U., {Meynet}, G., {et~al.} 2011, \nat, 472,
  454, \dodoi{10.1038/nature10000}

\bibitem[{{Chieffi} \& {Limongi}(2004)}]{chieffi2004}
{Chieffi}, A., \& {Limongi}, M. 2004, \apj, 608, 405, \dodoi{10.1086/392523}

\bibitem[{{Chieffi} \& {Limongi}(2013)}]{Chieffi2013}
---. 2013, \apj, 764, 21, \dodoi{10.1088/0004-637X/764/1/21}

\bibitem[{{Clausen} {et~al.}(2015){Clausen}, {Piro}, \& {Ott}}]{clausen2015}
{Clausen}, D., {Piro}, A.~L., \& {Ott}, C.~D. 2015, \apj, 799, 190,
  \dodoi{10.1088/0004-637X/799/2/190}

\bibitem[{{Collet} {et~al.}(2007){Collet}, {Asplund}, \&
  {Trampedach}}]{collet2007}
{Collet}, R., {Asplund}, M., \& {Trampedach}, R. 2007, \aap, 469, 687,
  \dodoi{10.1051/0004-6361:20066321}

\bibitem[{{De Silva} {et~al.}(2015){De Silva}, {Freeman}, {Bland-Hawthorn},
  {Martell}, {de Boer}, {Asplund}, {Keller}, {Sharma}, {Zucker}, {Zwitter},
  {Anguiano}, {Bacigalupo}, {Bayliss}, {Beavis}, {Bergemann}, {Campbell},
  {Cannon}, {Carollo}, {Casagrande}, {Casey}, {Da Costa}, {D'Orazi}, {Dotter},
  {Duong}, {Heger}, {Ireland}, {Kafle}, {Kos}, {Lattanzio}, {Lewis}, {Lin},
  {Lind}, {Munari}, {Nataf}, {O'Toole}, {Parker}, {Reid}, {Schlesinger},
  {Sheinis}, {Simpson}, {Stello}, {Ting}, {Traven}, {Watson}, {Wittenmyer},
  {Yong}, \& {{\v{Z}}erjal}}]{desilva2015}
{De Silva}, G.~M., {Freeman}, K.~C., {Bland-Hawthorn}, J., {et~al.} 2015,
  \mnras, 449, 2604, \dodoi{10.1093/mnras/stv327}

\bibitem[{{deBoer} {et~al.}(2017){deBoer}, {G{\"o}rres}, {Wiescher}, {Azuma},
  {Best}, {Brune}, {Fields}, {Jones}, {Pignatari}, {Sayre}, {Smith}, {Timmes},
  \& {Uberseder}}]{deBoer17}
{deBoer}, R.~J., {G{\"o}rres}, J., {Wiescher}, M., {et~al.} 2017, Reviews of
  Modern Physics, 89, 035007, \dodoi{10.1103/RevModPhys.89.035007}

\bibitem[{{Ebinger} {et~al.}(2019){Ebinger}, {Curtis}, {Fr{\"o}hlich},
  {Hempel}, {Perego}, {Liebend{\"o}rfer}, \& {Thielemann}}]{ebinger2019}
{Ebinger}, K., {Curtis}, S., {Fr{\"o}hlich}, C., {et~al.} 2019, \apj, 870, 1,
  \dodoi{10.3847/1538-4357/aae7c9}

\bibitem[{{Eisenstein} {et~al.}(2011){Eisenstein}, {Weinberg}, {Agol},
  {Aihara}, {Allende Prieto}, {Anderson}, {Arns}, {Aubourg}, {Bailey},
  {Balbinot}, {Barkhouser}, {Beers}, {Berlind}, {Bickerton}, {Bizyaev},
  {Blanton}, {Bochanski}, {Bolton}, {Bosman}, {Bovy}, {Brandt}, {Breslauer},
  {Brewington}, {Brinkmann}, {Brown}, {Brownstein}, {Burger}, {Busca},
  {Campbell}, {Cargile}, {Carithers}, {Carlberg}, {Carr}, {Chang}, {Chen},
  {Chiappini}, {Comparat}, {Connolly}, {Cortes}, {Croft}, {Cunha}, {da Costa},
  {Davenport}, {Dawson}, {De Lee}, {Porto de Mello}, {de Simoni}, {Dean},
  {Dhital}, {Ealet}, {Ebelke}, {Edmondson}, {Eiting}, {Escoffier}, {Esposito},
  {Evans}, {Fan}, {Femen{\'\i}a Castell{\'a}}, {Dutra Ferreira}, {Fitzgerald},
  {Fleming}, {Font-Ribera}, {Ford}, {Frinchaboy}, {Garc{\'\i}a P{\'e}rez},
  {Gaudi}, {Ge}, {Ghezzi}, {Gillespie}, {Gilmore}, {Girardi}, {Gott}, {Gould},
  {Grebel}, {Gunn}, {Hamilton}, {Harding}, {Harris}, {Hawley}, {Hearty},
  {Hennawi}, {Gonz{\'a}lez Hern{\'a}ndez}, {Ho}, {Hogg}, {Holtzman},
  {Honscheid}, {Inada}, {Ivans}, {Jiang}, {Jiang}, {Johnson}, {Jordan},
  {Jordan}, {Kauffmann}, {Kazin}, {Kirkby}, {Klaene}, {Knapp}, {Kneib},
  {Kochanek}, {Koesterke}, {Kollmeier}, {Kron}, {Lampeitl}, {Lang}, {Lawler},
  {Le Goff}, {Lee}, {Lee}, {Leisenring}, {Lin}, {Liu}, {Long}, {Loomis},
  {Lucatello}, {Lundgren}, {Lupton}, {Ma}, {Ma}, {MacDonald}, {Mack},
  {Mahadevan}, {Maia}, {Majewski}, {Makler}, {Malanushenko}, {Malanushenko},
  {Mandelbaum}, {Maraston}, {Margala}, {Maseman}, {Masters}, {McBride},
  {McDonald}, {McGreer}, {McMahon}, {Mena Requejo}, {M{\'e}nard},
  {Miralda-Escud{\'e}}, {Morrison}, {Mullally}, {Muna}, {Murayama}, {Myers},
  {Naugle}, {Neto}, {Nguyen}, {Nichol}, {Nidever}, {O'Connell}, {Ogando},
  {Olmstead}, {Oravetz}, {Padmanabhan}, {Paegert}, {Palanque-Delabrouille},
  {Pan}, {Pandey}, {Parejko}, {P{\^a}ris}, {Pellegrini}, {Pepper}, {Percival},
  {Petitjean}, {Pfaffenberger}, {Pforr}, {Phleps}, {Pichon}, {Pieri}, {Prada},
  {Price-Whelan}, {Raddick}, {Ramos}, {Reid}, {Reyle}, {Rich}, {Richards},
  {Rieke}, {Rieke}, {Rix}, {Robin}, {Rocha-Pinto}, {Rockosi}, {Roe},
  {Rollinde}, {Ross}, {Ross}, {Rossetto}, {S{\'a}nchez}, {Santiago}, {Sayres},
  {Schiavon}, {Schlegel}, {Schlesinger}, {Schmidt}, {Schneider}, {Sellgren},
  {Shelden}, {Sheldon}, {Shetrone}, {Shu}, {Silverman}, {Simmerer}, {Simmons},
  {Sivarani}, {Skrutskie}, {Slosar}, {Smee}, {Smith}, {Snedden}, {Stassun},
  {Steele}, {Steinmetz}, {Stockett}, {Stollberg}, {Strauss}, {Szalay},
  {Tanaka}, {Thakar}, {Thomas}, {Tinker}, {Tofflemire}, {Tojeiro}, {Tremonti},
  {Vargas Maga{\~n}a}, {Verde}, {Vogt}, {Wake}, {Wan}, {Wang}, {Weaver},
  {White}, {White}, {Wilson}, {Wisniewski}, {Wood-Vasey}, {Yanny}, {Yasuda},
  {Y{\`e}che}, {York}, {Young}, {Zasowski}, {Zehavi}, \&
  {Zhao}}]{eisenstein2011}
{Eisenstein}, D.~J., {Weinberg}, D.~H., {Agol}, E., {et~al.} 2011, \aj, 142,
  72, \dodoi{10.1088/0004-6256/142/3/72}

\bibitem[{{Ertl} {et~al.}(2016){Ertl}, {Janka}, {Woosley}, {Sukhbold}, \&
  {Ugliano}}]{ertl2016}
{Ertl}, T., {Janka}, H.~T., {Woosley}, S.~E., {Sukhbold}, T., \& {Ugliano}, M.
  2016, \apj, 818, 124, \dodoi{10.3847/0004-637X/818/2/124}

\bibitem[{{Ertl} {et~al.}(2020){Ertl}, {Woosley}, {Sukhbold}, \&
  {Janka}}]{ertl2020}
{Ertl}, T., {Woosley}, S.~E., {Sukhbold}, T., \& {Janka}, H.~T. 2020, \apj,
  890, 51, \dodoi{10.3847/1538-4357/ab6458}

\bibitem[{{Farmer} {et~al.}(2020){Farmer}, {Renzo}, {de Mink}, {Fishbach}, \&
  {Justham}}]{farmer2020}
{Farmer}, R., {Renzo}, M., {de Mink}, S.~E., {Fishbach}, M., \& {Justham}, S.
  2020, \apjl, 902, L36, \dodoi{10.3847/2041-8213/abbadd}

\bibitem[{{Farmer} {et~al.}(2019){Farmer}, {Renzo}, {de Mink}, {Marchant}, \&
  {Justham}}]{farmer2019}
{Farmer}, R., {Renzo}, M., {de Mink}, S.~E., {Marchant}, P., \& {Justham}, S.
  2019, \apj, 887, 53, \dodoi{10.3847/1538-4357/ab518b}

\bibitem[{{Finlator} \& {Dav{\'e}}(2008)}]{Finlator2008}
{Finlator}, K., \& {Dav{\'e}}, R. 2008, \mnras, 385, 2181,
  \dodoi{10.1111/j.1365-2966.2008.12991.x}

\bibitem[{{Frischknecht} {et~al.}(2012){Frischknecht}, {Hirschi}, \&
  {Thielemann}}]{frischknecht2012}
{Frischknecht}, U., {Hirschi}, R., \& {Thielemann}, F.~K. 2012, \aap, 538, L2,
  \dodoi{10.1051/0004-6361/201117794}

\bibitem[{{Grevesse} \& {Sauval}(1998)}]{grevesse1998}
{Grevesse}, N., \& {Sauval}, A.~J. 1998, \ssr, 85, 161,
  \dodoi{10.1023/A:1005161325181}

\bibitem[{{Griffith} {et~al.}(2019){Griffith}, {Johnson}, \&
  {Weinberg}}]{griffith19}
{Griffith}, E., {Johnson}, J.~A., \& {Weinberg}, D.~H. 2019, \apj, 886, 84,
  \dodoi{10.3847/1538-4357/ab4b5d}

\bibitem[{{Griffith} {et~al.}(2021){Griffith}, {Weinberg}, {Johnson}, {Beaton},
  {Garc{\'\i}a-Hern{\'a}ndez}, {Hasselquist}, {Holtzman}, {Johnson},
  {J{\"o}nsson}, {Lane}, {Nataf}, \& {Roman-Lopes}}]{Griffith2021}
{Griffith}, E., {Weinberg}, D.~H., {Johnson}, J.~A., {et~al.} 2021, \apj, 909,
  77, \dodoi{10.3847/1538-4357/abd6be}

\bibitem[{{Harris} {et~al.}(2020){Harris}, {Millman}, {van der Walt},
  {Gommers}, {Virtanen}, {Cournapeau}, {Wieser}, {Taylor}, {Berg}, {Smith},
  {Kern}, {Picus}, {Hoyer}, {van Kerkwijk}, {Brett}, {Haldane}, {del R{\'\i}o},
  {Wiebe}, {Peterson}, {G{\'e}rard-Marchant}, {Sheppard}, {Reddy}, {Weckesser},
  {Abbasi}, {Gohlke}, \& {Oliphant}}]{harris2020}
{Harris}, C.~R., {Millman}, K.~J., {van der Walt}, S.~J., {et~al.} 2020, \nat,
  585, 357, \dodoi{10.1038/s41586-020-2649-2}

\bibitem[{{Hayek} {et~al.}(2011){Hayek}, {Asplund}, {Collet}, \&
  {Nordlund}}]{hayek2011}
{Hayek}, W., {Asplund}, M., {Collet}, R., \& {Nordlund}, {\r{A}}. 2011, \aap,
  529, A158, \dodoi{10.1051/0004-6361/201015782}

\bibitem[{{Heger} {et~al.}(2003){Heger}, {Fryer}, {Woosley}, {Langer}, \&
  {Hartmann}}]{heger2003}
{Heger}, A., {Fryer}, C.~L., {Woosley}, S.~E., {Langer}, N., \& {Hartmann},
  D.~H. 2003, \apj, 591, 288, \dodoi{10.1086/375341}

\bibitem[{{Horiuchi} {et~al.}(2014){Horiuchi}, {Nakamura}, {Takiwaki},
  {Kotake}, \& {Tanaka}}]{horiuchi2014}
{Horiuchi}, S., {Nakamura}, K., {Takiwaki}, T., {Kotake}, K., \& {Tanaka}, M.
  2014, \mnras, 445, L99, \dodoi{10.1093/mnrasl/slu146}

\bibitem[{{Hunter}(2007)}]{hunter2007}
{Hunter}, J.~D. 2007, Computing in Science and Engineering, 9, 90,
  \dodoi{10.1109/MCSE.2007.55}

\bibitem[{{Janka} {et~al.}(2016){Janka}, {Melson}, \& {Summa}}]{janka2016}
{Janka}, H.-T., {Melson}, T., \& {Summa}, A. 2016, Annual Review of Nuclear and
  Particle Science, 66, 341, \dodoi{10.1146/annurev-nucl-102115-044747}

\bibitem[{{Jerkstrand} {et~al.}(2015){Jerkstrand}, {Ergon}, {Smartt},
  {Fransson}, {Sollerman}, {Taubenberger}, {Bersten}, \&
  {Spyromilio}}]{jerkstrand2015}
{Jerkstrand}, A., {Ergon}, M., {Smartt}, S.~J., {et~al.} 2015, \aap, 573, A12,
  \dodoi{10.1051/0004-6361/201423983}

\bibitem[{{Johnson} \& {Weinberg}(2020)}]{johnson2020}
{Johnson}, J.~W., \& {Weinberg}, D.~H. 2020, \mnras, 498, 1364,
  \dodoi{10.1093/mnras/staa2431}

\bibitem[{{Kib{\'e}di} {et~al.}(2020){Kib{\'e}di}, {Alshahrani}, {Stuchbery},
  {Larsen}, {G{\"o}rgen}, {Siem}, {Guttormsen}, {Giacoppo}, {Morales}, {Sahin},
  {Tveten}, {Garrote}, {Campo}, {Eriksen}, {Klintefjord}, {Maharramova},
  {Nyhus}, {Tornyi}, {Renstr{\o}m}, \& {Paulsen}}]{Kib20}
{Kib{\'e}di}, T., {Alshahrani}, B., {Stuchbery}, A.~E., {et~al.} 2020, \prl,
  125, 182701, \dodoi{10.1103/PhysRevLett.125.182701}

\bibitem[{{Kiselman}(1993)}]{kiselman1993}
{Kiselman}, D. 1993, \aap, 275, 269

\bibitem[{{Kroupa}(2001)}]{kroupa2001}
{Kroupa}, P. 2001, \mnras, 322, 231, \dodoi{10.1046/j.1365-8711.2001.04022.x}

\bibitem[{{Kroupa} {et~al.}(1993){Kroupa}, {Tout}, \& {Gilmore}}]{Kroupa1993}
{Kroupa}, P., {Tout}, C.~A., \& {Gilmore}, G. 1993, \mnras, 262, 545,
  \dodoi{10.1093/mnras/262.3.545}

\bibitem[{{Laplace} {et~al.}(2021){Laplace}, {Justham}, {Renzo}, {G{\"o}tberg},
  {Farmer}, {Vartanyan}, \& {de Mink}}]{Laplace2021}
{Laplace}, E., {Justham}, S., {Renzo}, M., {et~al.} 2021, arXiv e-prints,
  arXiv:2102.05036.
\newblock \doarXiv{2102.05036}

\bibitem[{{Limongi} \& {Chieffi}(2006)}]{limongi2016}
{Limongi}, M., \& {Chieffi}, A. 2006, \apj, 647, 483, \dodoi{10.1086/505164}

\bibitem[{{Limongi} \& {Chieffi}(2018)}]{lc18}
---. 2018, \apjs, 237, 13, \dodoi{10.3847/1538-4365/aacb24}

\bibitem[{{Lodders}(2003)}]{lodders2003}
{Lodders}, K. 2003, \apj, 591, 1220, \dodoi{10.1086/375492}

\bibitem[{{Lodders}(2010)}]{lodders2010}
---. 2010, Astrophysics and Space Science Proceedings, 16, 379,
  \dodoi{10.1007/978-3-642-10352-0_8}

\bibitem[{{Lovegrove} \& {Woosley}(2013)}]{lovegrove2013}
{Lovegrove}, E., \& {Woosley}, S.~E. 2013, \apj, 769, 109,
  \dodoi{10.1088/0004-637X/769/2/109}

\bibitem[{{Mabanta} {et~al.}(2019){Mabanta}, {Murphy}, \&
  {Dolence}}]{mabanta2019}
{Mabanta}, Q.~A., {Murphy}, J.~W., \& {Dolence}, J.~C. 2019, \apj, 887, 43,
  \dodoi{10.3847/1538-4357/ab4bcc}

\bibitem[{{Majewski} {et~al.}(2017){Majewski}, {Schiavon}, {Frinchaboy},
  {Allende Prieto}, {Barkhouser}, {Bizyaev}, {Blank}, {Brunner}, {Burton},
  {Carrera}, {Chojnowski}, {Cunha}, {Epstein}, {Fitzgerald}, {Garc{\'\i}a
  P{\'e}rez}, {Hearty}, {Henderson}, {Holtzman}, {Johnson}, {Lam}, {Lawler},
  {Maseman}, {M{\'e}sz{\'a}ros}, {Nelson}, {Nguyen}, {Nidever}, {Pinsonneault},
  {Shetrone}, {Smee}, {Smith}, {Stolberg}, {Skrutskie}, {Walker}, {Wilson},
  {Zasowski}, {Anders}, {Basu}, {Beland}, {Blanton}, {Bovy}, {Brownstein},
  {Carlberg}, {Chaplin}, {Chiappini}, {Eisenstein}, {Elsworth}, {Feuillet},
  {Fleming}, {Galbraith-Frew}, {Garc{\'\i}a}, {Garc{\'\i}a-Hern{\'a}ndez},
  {Gillespie}, {Girardi}, {Gunn}, {Hasselquist}, {Hayden}, {Hekker}, {Ivans},
  {Kinemuchi}, {Klaene}, {Mahadevan}, {Mathur}, {Mosser}, {Muna}, {Munn},
  {Nichol}, {O'Connell}, {Parejko}, {Robin}, {Rocha-Pinto}, {Schultheis},
  {Serenelli}, {Shane}, {Silva Aguirre}, {Sobeck}, {Thompson}, {Troup},
  {Weinberg}, \& {Zamora}}]{majewski17}
{Majewski}, S.~R., {Schiavon}, R.~P., {Frinchaboy}, P.~M., {et~al.} 2017, \aj,
  154, 94, \dodoi{10.3847/1538-3881/aa784d}

\bibitem[{{Martell} {et~al.}(2017){Martell}, {Sharma}, {Buder}, {Duong},
  {Schlesinger}, {Simpson}, {Lind}, {Ness}, {Marshall}, {Asplund},
  {Bland-Hawthorn}, {Casey}, {De Silva}, {Freeman}, {Kos}, {Lin}, {Zucker},
  {Zwitter}, {Anguiano}, {Bacigalupo}, {Carollo}, {Casagrande}, {Da Costa},
  {Horner}, {Huber}, {Hyde}, {Kafle}, {Lewis}, {Nataf}, {Navin}, {Stello},
  {Tinney}, {Watson}, \& {Wittenmyer}}]{martell2017}
{Martell}, S.~L., {Sharma}, S., {Buder}, S., {et~al.} 2017, \mnras, 465, 3203,
  \dodoi{10.1093/mnras/stw2835}

\bibitem[{{Matteucci} \& {Francois}(1989)}]{matteucci1989}
{Matteucci}, F., \& {Francois}, P. 1989, \mnras, 239, 885,
  \dodoi{10.1093/mnras/239.3.885}

\bibitem[{{Mishenina} {et~al.}(2019){Mishenina}, {Pignatari}, {Gorbaneva},
  {Bisterzo}, {Travaglio}, {Thielemann}, \& {Soubiran}}]{mishenina2019}
{Mishenina}, T., {Pignatari}, M., {Gorbaneva}, T., {et~al.} 2019, \mnras, 484,
  3846, \dodoi{10.1093/mnras/stz178}

\bibitem[{{M{\"u}ller}(2020)}]{bernhard2020}
{M{\"u}ller}, B. 2020, Living Reviews in Computational Astrophysics, 6, 3,
  \dodoi{10.1007/s41115-020-0008-5}

\bibitem[{{M{\"u}ller} {et~al.}(2016){M{\"u}ller}, {Heger}, {Liptai}, \&
  {Cameron}}]{muller2016}
{M{\"u}ller}, B., {Heger}, A., {Liptai}, D., \& {Cameron}, J.~B. 2016, \mnras,
  460, 742, \dodoi{10.1093/mnras/stw1083}

\bibitem[{{O'Connor} \& {Ott}(2011)}]{oconnor2011}
{O'Connor}, E., \& {Ott}, C.~D. 2011, \apj, 730, 70,
  \dodoi{10.1088/0004-637X/730/2/70}

\bibitem[{{Palla} {et~al.}(2020){Palla}, {Matteucci}, {Spitoni}, {Vincenzo}, \&
  {Grisoni}}]{palla2020}
{Palla}, M., {Matteucci}, F., {Spitoni}, E., {Vincenzo}, F., \& {Grisoni}, V.
  2020, \mnras, 498, 1710, \dodoi{10.1093/mnras/staa2437}

\bibitem[{{Patton} \& {Sukhbold}(2020)}]{Patton2020}
{Patton}, R.~A., \& {Sukhbold}, T. 2020, \mnras, 499, 2803,
  \dodoi{10.1093/mnras/staa3029}

\bibitem[{{Peeples} \& {Shankar}(2011)}]{Peeples2011}
{Peeples}, M.~S., \& {Shankar}, F. 2011, \mnras, 417, 2962,
  \dodoi{10.1111/j.1365-2966.2011.19456.x}

\bibitem[{{Pejcha} \& {Thompson}(2015)}]{pejcha2015}
{Pejcha}, O., \& {Thompson}, T.~A. 2015, \apj, 801, 90,
  \dodoi{10.1088/0004-637X/801/2/90}

\bibitem[{{Pignatari} {et~al.}(2008){Pignatari}, {Gallino}, {Meynet},
  {Hirschi}, {Herwig}, \& {Wiescher}}]{pignatari2008}
{Pignatari}, M., {Gallino}, R., {Meynet}, G., {et~al.} 2008, \apjl, 687, L95,
  \dodoi{10.1086/593350}

\bibitem[{{Prantzos} {et~al.}(2018){Prantzos}, {Abia}, {Limongi}, {Chieffi}, \&
  {Cristallo}}]{prantzos2018}
{Prantzos}, N., {Abia}, C., {Limongi}, M., {Chieffi}, A., \& {Cristallo}, S.
  2018, \mnras, 476, 3432, \dodoi{10.1093/mnras/sty316}

\bibitem[{{Raithel} {et~al.}(2018){Raithel}, {Sukhbold}, \&
  {{\"O}zel}}]{raithel2018}
{Raithel}, C.~A., {Sukhbold}, T., \& {{\"O}zel}, F. 2018, \apj, 856, 35,
  \dodoi{10.3847/1538-4357/aab09b}

\bibitem[{{Rauscher} {et~al.}(2002){Rauscher}, {Heger}, {Hoffman}, \&
  {Woosley}}]{rauscher2002}
{Rauscher}, T., {Heger}, A., {Hoffman}, R.~D., \& {Woosley}, S.~E. 2002, \apj,
  576, 323, \dodoi{10.1086/341728}

\bibitem[{{Remillard} \& {McClintock}(2006)}]{remillard2006}
{Remillard}, R.~A., \& {McClintock}, J.~E. 2006, \araa, 44, 49,
  \dodoi{10.1146/annurev.astro.44.051905.092532}

\bibitem[{{Ritter} {et~al.}(2018){Ritter}, {Andrassy}, {C{\^o}t{\'e}},
  {Herwig}, {Woodward}, {Pignatari}, \& {Jones}}]{ritter2018}
{Ritter}, C., {Andrassy}, R., {C{\^o}t{\'e}}, B., {et~al.} 2018, \mnras, 474,
  L1, \dodoi{10.1093/mnrasl/slx126}

\bibitem[{{Romano} {et~al.}(2005){Romano}, {Chiappini}, {Matteucci}, \&
  {Tosi}}]{romano2005}
{Romano}, D., {Chiappini}, C., {Matteucci}, F., \& {Tosi}, M. 2005, \aap, 430,
  491, \dodoi{10.1051/0004-6361:20048222}

\bibitem[{{Romano} {et~al.}(2010){Romano}, {Karakas}, {Tosi}, \&
  {Matteucci}}]{romano2010}
{Romano}, D., {Karakas}, A.~I., {Tosi}, M., \& {Matteucci}, F. 2010, \aap, 522,
  A32, \dodoi{10.1051/0004-6361/201014483}

\bibitem[{{Rybizki} {et~al.}(2017){Rybizki}, {Just}, \& {Rix}}]{rybizki2017}
{Rybizki}, J., {Just}, A., \& {Rix}, H.-W. 2017, \aap, 605, A59,
  \dodoi{10.1051/0004-6361/201730522}

\bibitem[{{Salpeter}(1955)}]{Salpeter1955}
{Salpeter}, E.~E. 1955, \apj, 121, 161, \dodoi{10.1086/145971}

\bibitem[{{Sander} {et~al.}(2020){Sander}, {Vink}, \& {Hamann}}]{sander2020}
{Sander}, A. A.~C., {Vink}, J.~S., \& {Hamann}, W.~R. 2020, \mnras, 491, 4406,
  \dodoi{10.1093/mnras/stz3064}

\bibitem[{{Scalo}(1986)}]{Scalo1986}
{Scalo}, J.~M. 1986, \fcp, 11, 1

\bibitem[{{Smartt}(2015)}]{smartt2015}
{Smartt}, S.~J. 2015, \pasa, 32, e016, \dodoi{10.1017/pasa.2015.17}

\bibitem[{{Spitoni} {et~al.}(2021){Spitoni}, {Verma}, {Silva Aguirre},
  {Vincenzo}, {Matteucci}, {Vai{\v{c}}ekauskait{\.{e}}}, {Palla}, {Grisoni}, \&
  {Calura}}]{spitoni2021}
{Spitoni}, E., {Verma}, K., {Silva Aguirre}, V., {et~al.} 2021, \aap, 647, A73,
  \dodoi{10.1051/0004-6361/202039864}

\bibitem[{{Sukhbold} {et~al.}(2016){Sukhbold}, {Ertl}, {Woosley}, {Brown}, \&
  {Janka}}]{sukhbold2016}
{Sukhbold}, T., {Ertl}, T., {Woosley}, S.~E., {Brown}, J.~M., \& {Janka}, H.~T.
  2016, \apj, 821, 38, \dodoi{10.3847/0004-637X/821/1/38}

\bibitem[{{Sukhbold} \& {Woosley}(2014)}]{sukhbold2014}
{Sukhbold}, T., \& {Woosley}, S.~E. 2014, \apj, 783, 10,
  \dodoi{10.1088/0004-637X/783/1/10}

\bibitem[{{Sukhbold} {et~al.}(2018){Sukhbold}, {Woosley}, \&
  {Heger}}]{sukhbold2018}
{Sukhbold}, T., {Woosley}, S.~E., \& {Heger}, A. 2018, \apj, 860, 93,
  \dodoi{10.3847/1538-4357/aac2da}

\bibitem[{{Tur} {et~al.}(2007){Tur}, {Heger}, \& {Austin}}]{tur2007}
{Tur}, C., {Heger}, A., \& {Austin}, S.~M. 2007, \apj, 671, 821,
  \dodoi{10.1086/523095}

\bibitem[{{Ugliano} {et~al.}(2012){Ugliano}, {Janka}, {Marek}, \&
  {Arcones}}]{ugliano2012}
{Ugliano}, M., {Janka}, H.-T., {Marek}, A., \& {Arcones}, A. 2012, \apj, 757,
  69, \dodoi{10.1088/0004-637X/757/1/69}

\bibitem[{{Villante} {et~al.}(2014){Villante}, {Serenelli}, {Delahaye}, \&
  {Pinsonneault}}]{villante2014}
{Villante}, F.~L., {Serenelli}, A.~M., {Delahaye}, F., \& {Pinsonneault}, M.~H.
  2014, \apj, 787, 13, \dodoi{10.1088/0004-637X/787/1/13}

\bibitem[{{Vincenzo} {et~al.}(2021){Vincenzo}, {Thompson}, {Weinberg},
  {Griffith}, {Johnson}, \& {Johnson}}]{Vincenzo2021}
{Vincenzo}, F., {Thompson}, T.~A., {Weinberg}, D.~H., {et~al.} 2021, arXiv
  e-prints, arXiv:2102.04920.
\newblock \doarXiv{2102.04920}

\bibitem[{{Vink}(2017)}]{vink2017}
{Vink}, J.~S. 2017, \aap, 607, L8, \dodoi{10.1051/0004-6361/201731902}

\bibitem[{{Vlasov} {et~al.}(2017){Vlasov}, {Metzger}, {Lippuner}, {Roberts}, \&
  {Thompson}}]{Vlasov2017}
{Vlasov}, A.~D., {Metzger}, B.~D., {Lippuner}, J., {Roberts}, L.~F., \&
  {Thompson}, T.~A. 2017, \mnras, 468, 1522, \dodoi{10.1093/mnras/stx478}

\bibitem[{{Weaver} {et~al.}(1978){Weaver}, {Zimmerman}, \&
  {Woosley}}]{weaver1978}
{Weaver}, T.~A., {Zimmerman}, G.~B., \& {Woosley}, S.~E. 1978, \apj, 225, 1021,
  \dodoi{10.1086/156569}

\bibitem[{{Weinberg} {et~al.}(2017){Weinberg}, {Andrews}, \&
  {Freudenburg}}]{Weinberg2017}
{Weinberg}, D.~H., {Andrews}, B.~H., \& {Freudenburg}, J. 2017, \apj, 837, 183,
  \dodoi{10.3847/1538-4357/837/2/183}

\bibitem[{{Weinberg} {et~al.}(2019){Weinberg}, {Holtzman}, {Hasselquist},
  {Bird}, {Johnson}, {Shetrone}, {Sobeck}, {Allende Prieto}, {Bizyaev},
  {Carrera}, {Cohen}, {Cunha}, {Ebelke}, {Fernandez-Trincado},
  {Garc{\'\i}a-Hern{\'a}ndez}, {Hayes}, {J{\"o}nsson}, {Lane}, {Majewski},
  {Malanushenko}, {M{\'e}sz{\'a}ros}, {Nidever}, {Nitschelm}, {Pan}, {Rix},
  {Rybizki}, {Schiavon}, {Schneider}, {Wilson}, \& {Zamora}}]{weinberg2019}
{Weinberg}, D.~H., {Holtzman}, J.~A., {Hasselquist}, S., {et~al.} 2019, ApJ,
  874, 102, \dodoi{10.3847/1538-4357/ab07c7}

\bibitem[{{Woosley} \& {Heger}(2007)}]{woosley2007}
{Woosley}, S.~E., \& {Heger}, A. 2007, \physrep, 442, 269,
  \dodoi{10.1016/j.physrep.2007.02.009}

\bibitem[{{Woosley} \& {Heger}(2015)}]{woosley2015}
---. 2015, \apj, 810, 34, \dodoi{10.1088/0004-637X/810/1/34}

\bibitem[{{Woosley} \& {Heger}(2021)}]{woosley2021}
---. 2021, arXiv e-prints, arXiv:2103.07933.
\newblock \doarXiv{2103.07933}

\bibitem[{{Woosley} {et~al.}(2002){Woosley}, {Heger}, \&
  {Weaver}}]{woosley2002}
{Woosley}, S.~E., {Heger}, A., \& {Weaver}, T.~A. 2002, Reviews of Modern
  Physics, 74, 1015, \dodoi{10.1103/RevModPhys.74.1015}

\bibitem[{{Woosley} \& {Weaver}(1995)}]{woosley1995}
{Woosley}, S.~E., \& {Weaver}, T.~A. 1995, \apjs, 101, 181,
  \dodoi{10.1086/192237}

\bibitem[{{Yoon}(2017)}]{yoon2017}
{Yoon}, S.-C. 2017, \mnras, 470, 3970, \dodoi{10.1093/mnras/stx1496}

\bibitem[{{Zahid} {et~al.}(2012){Zahid}, {Dima}, {Kewley}, {Erb}, \&
  {Dav{\'e}}}]{Zahid2012}
{Zahid}, H.~J., {Dima}, G.~I., {Kewley}, L.~J., {Erb}, D.~K., \& {Dav{\'e}}, R.
  2012, \apj, 757, 54, \dodoi{10.1088/0004-637X/757/1/54}

\bibitem[{{Zhao} {et~al.}(2016){Zhao}, {Mashonkina}, {Yan}, {Alexeeva},
  {Kobayashi}, {Pakhomov}, {Shi}, {Sitnova}, {Tan}, {Zhang}, {Zhang}, {Zhou},
  {Bolte}, {Chen}, {Li}, {Liu}, \& {Zhai}}]{zhao2016}
{Zhao}, G., {Mashonkina}, L., {Yan}, H.~L., {et~al.} 2016, \apj, 833, 225,
  \dodoi{10.3847/1538-4357/833/2/225}

\end{thebibliography}
\bibliographystyle{aasjournal}

\end{document}